\documentclass[%
 reprint,
 amsmath,amssymb,
 aps,
]{revtex4-1}

\usepackage{blindtext}
\usepackage{booktabs}
\usepackage{graphicx}
\usepackage{bm}

\usepackage[utf8]{inputenc}
\usepackage{amsmath,color}
\usepackage{amsfonts}
\usepackage{amssymb}
\usepackage{graphicx}
\usepackage{soul}
\usepackage{parskip}
\usepackage{natbib}
\usepackage{algorithm}
\usepackage{algpseudocode}
\usepackage{blindtext}
\usepackage{booktabs}
\usepackage{graphicx}

\begin{document}


\title{Self-falsifiable Hierarchical Detection of Overlapping Communities On Social Networks}

\author{Tianyi Li}
\thanks{Corresponding author: tianyil@mit.edu}
\affiliation{System Dynamics Group, Sloan School of Management, Massachusetts Institute of Technology}
\author{Pan Zhang}%
\affiliation{Key Laboratory of Theoretical Physics, Institute of Theoretical Physics,Chinese Academy of Sciences}

\date{\today}

\begin{abstract}
No community detection algorithm can be optimal for all possible networks, thus it is important to identify whether the algorithm is suitable for a given network. We propose a multi-step algorithmic solution scheme for overlapping community detection based on an advanced label propagation process, which imitates the community formation process on social networks. Our algorithm is parameter-free and is able to reveal the hierarchical order of communities in the graph. The unique property of our solution scheme is \textit{self-falsifiability}; an automatic quality check of the results is conducted after the detection, and the fitness of the algorithm for the specific network is reported. Extensive experiments show that our algorithm is self-consistent, reliable on networks of a wide range of size and different sorts, and is more robust than existing algorithms on both sparse and large-scale social networks. Results further suggest that our solution scheme may uncover features of networks' intrinsic community structures. 
\end{abstract}

\maketitle

Community detection is a central topic in network science. Pioneered by works represented by \citet{Pet2005}, in recent years more and more studies focus on the detection of overlapping communities as opposed to exhaustive communities, a division also regarded as soft-partitioning versus hard-partitioning. Traditional detection algorithms relying on the optimization of certain metrics, e.g. modularity, conductance etc., are often not directly applicable to overlapping communities, and therefore a lot of new tools are designed based on various ideas, including link communities \citep{Aet2010}, clique percolation \citep{Xet2013}, seed set expansion \citep{AL2006, Aet2006, Let2008, KK2014, Bet2017}, label propagation \citep{Ret2007, Xet2011,Cet2012, CR2017}, local spectral clustering method \citep{Let2018}, and methods based on statistical inference, such as Infomap \citep{RB2008}, the stochastic block model (SBM, \citep{KN2011,Set2011}; in particular, methods adopting the belief propagation algorithm \citep{ZM2014,Zet2016}), and other generative models \citep{Bet2011,Cet2017}. 

Despite the success of different solution schemes on various application fronts, some weak points of existing community detection algorithms could be pinned down in practice, which we believe might be problematic in certain cases (see Appendix A for a detailed discussion). These weaknesses include: (1) many solution schemes are over-parameterized, and in some cases the tuning of parameters depends largely on unwarranted heuristics; (2) many scalable methods based on the seed set expansion process \citep{C2005,AL2006,Aet2006,B2008,LLet2009,LLet2011,Ret2012,Let2013} may lack well-designed seeding strategies \citep{KK2014,Get2016,Bet2017} and often rely on ad-hoc strategies; (3) some algorithms that claim to be local, as opposed to methods based on an optimization over the entire graph, in fact still optimize on the community level and thus do not guarantee complete locality; (4) the number of communities in the graph is often pre-determined in certain algorithms, which might not be a good treatment, despite its claimed advantage \citep{FH2016} and the possible determination by the non-backtracking matrix \citep{Ket2013}; (5) the overlapping communities revealed by some algorithms are in fact still exhaustive in their corresponding link communities \citep{Aet2010}, which should not be an implicit constraint imposed by algorithms; (6) in many cases, the revealed communities do not follow any order and instead are treated as of equal significance to the graph (``blended'' \citep{FH2016}), which may deviate from realistic situations; (7) most algorithms assume that all nodes in the graph should belong to at least one community, without taking care of those isolated nodes that do not have any community membership \citep{Get2005,Set2010,Wet2013,Get2018}; (8) finally, a notification of the quality of detection results is not incorporated in most algorithms, failing to indicate the inevitable limited applicability of the method.

In a recent study, \citet{Pet2017} showed that, community detection is such an ill-defined problem that intrinsically no algorithm could be the optimal solution for all tasks, essentially a variant of the No-Free-Lunch theorem. Although this result seems to make the probe of community detection algorithms less meaningful, we argue that various streams of community detection ideas have embodied valuable beliefs for solving this problem and it is still useful to devise new approaches that inherit and combine the successful ideas of previous attempts. However, the most important lesson from \citep{Pet2017} is that, as noted by point (8) in the above discussion, when implemented on an arbitrary graph, a reliable community detection scheme should be able to indicate the extent of its applicability on the specific network, i.e., the extent of imperfectness of its detection results. We believe that this property of \textit{self-falsifiability} is an important missing piece in most existing algorithms.

Aiming at circumventing these weaknesses, correspondingly, we formulate our integrated belief for the overlapping community detection problem, with an emphasis on social networks where nodes represent human beings. This emphasis implies that the determination of community membership should incorporate  behavioral features, rather than being a completely mechanical process. We conclude important insights from multiple streams of existing algorithms \citep{FH2016} and integrate them into our belief; some extra attention to propagation-based approaches is paid, which are missing in \citep{FH2016} yet play a key role to account for the dynamic nature of social networks. The integrated belief consists of six aspects and is sketched in Figure 1.

(i) \textbf{Overlappedness}  One node could be able to belong to multiple communities, and its ``strength'' in different communities, e.g. in terms of the degree of attachment, should not be assumed to be homogeneous over its multiple memberships. Meanwhile, the corresponding link communities derived from the nodes' overlapping community assignment, should not be assumed to be exhaustive. One link could belong to different communities, such that the overlapping of communities should allow two communities to share a finite part of their components \citep[e.g.][]{Het2018}, consisting of both nodes and edges, as opposed to the case in \citet{Aet2010}.

(ii) \textbf{Different Roles of Nodes} Depending on the roles in communities, nodes in a typical social network may fall into five categories: hubs (sources), inner members of communities, boundaries (sinks), leaf nodes, and isolated nodes. Communities are initiated by hubs, but are finalized by sink nodes who set the boundaries, which are nodes that belong to more than one communities.  Edges are not natural boundaries of communities, as implied by (i). Isolated nodes belong to no communities; leaf nodes have only one neighbor and thus play a trivial role in the detection process.

(iii) \textbf{Behavioral Locality} In social networks, it is difficult for nodes to be acknowledged with information regarding the entire graph, even information regarding the other part of their communities. Therefore, in human networks the decision of community membership should be \textit{local}, following behavioral rules on nodes, instead of being derived from any optimization standpoint.

(iv) \textbf{Propagatory Formulation of Communities} On social networks, communities emerge along the propagation of information and action, hence methods imitating the propagation process have the advantage in revealing community structures. Many nodes could be the source (seeds) of the propagation, while some of them are dominated by others and only a few could be successfully identified as hubs. During the propagation, each node should be associated with a finite memory \citep{G2010}, recording the history of infections it receives from multiple communities. The determination of the community membership as well as the strength of the membership emerge from the infection history. 

(v) \textbf{Order of Communities} Communities on graphs should follow a hierarchical order \citep{Set2007,Cet2008, Set2008,LLet2009,LLet2011,P2014,Yet2017}: iteratively, the aggregation of small communities gives rise to bigger communities, and the entire graph is the single ultimate community.

(vi) \textbf{Self-Falsifiability} The applicability of any community detection algorithm is limited \citep{Pet2017}. When implemented on graphs with an arbitrary topology, detection algorithms should be able to quantitatively indicate the quality of the detection results, due to their varied applicability on specific graphs. In particular, a reliable detection algorithm is supposed to notify its potential failure on certain networks.

\begin{figure*}
\begin{minipage}{\linewidth}

\centering   
	\includegraphics[width=6in]{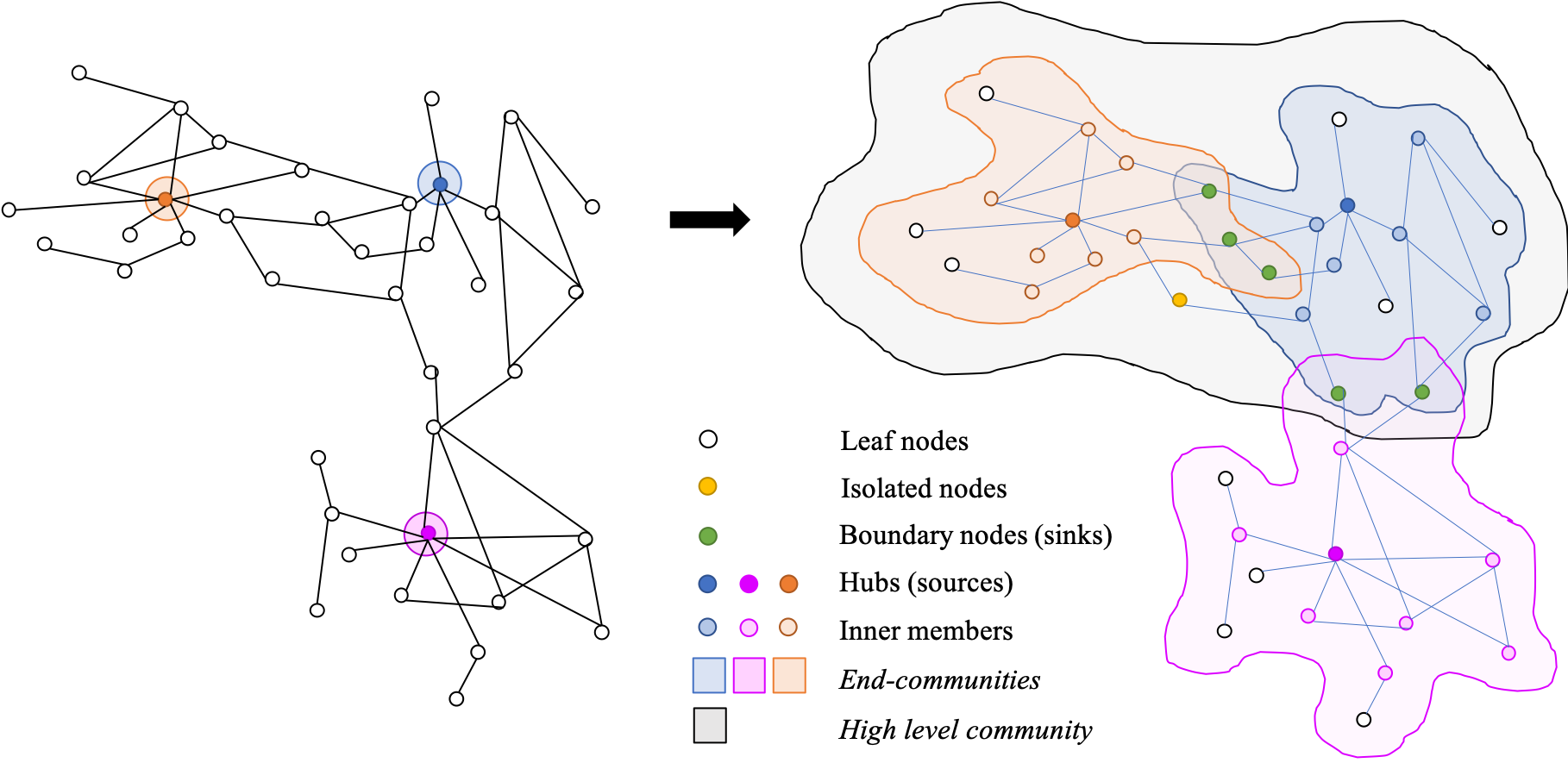}  
\caption{Illustration of the integrated belief underlying the proposed detection scheme.} 

\end{minipage}
\end{figure*}

Based on our integrated belief, we proposed a multi-step \citep{Let2008,Wet2013} algorithmic solution scheme for overlapping community detection. Our approach is in line with the DBSCAN algorithm \citep{Eet1996, Cet2013,Set2017} and the SHRINK algorithm \citep{Set2010}, but having a more transparent and better quantified workflow, along with two new features: \textit{parameter-free}, and \textit{self-falsifiable}. The framework consists of four steps. First, nodes are identified with different roles in the graph based on their centrality scores, among which nodes having local centrality peaks are detected as the hubs (sources) of end-communities. Next, a diffusive label propagation process is initiated, starting simultaneously from all hubs, and spread on the entire graph. The determination (expansion) of end-communities converges at the end of the propagation process. Third, the distance matrix of end-communities is calculated, which facilitates the construction of the community hierarchy by aggregating end-communities in an upward fashion along the distance matrix. The entire graph becomes the ultimate community on the top of the hierarchy. In the end, the quality of the obtained community hierarchy is automatically checked and quantitatively indicated after the detection, and suggestions for the cutoff levels of the community hierarchy are provided. Details of the four steps of the detection scheme are discussed in the next section. Performance of the algorithm is studied and discussed in the following sections.


\section*{Multi-step Detection Algorithm}

\textit{Step1}: \textbf{Identification of nodes' roles}. Assume a graph with $N$ nodes and $E$ edges. Given the connectivity matrix $A = \{a_{ij}\}$ of the graph, first we calculate the centrality scores $c_i$ of each node $i$, and find the set of nodes whose centrality score is local peak, i.e. whose centrality is \textit{no less than} all its neighbors. In theory, different kinds of centrality measures could be used. Path-based centralities such as betweenness centrality or closeness centrality may not be suitable for our current setting; density-based measures such as degree centrality and eigenvector centrality are more appropriate to apply. Among these nodes that are local peaks, those whose centrality is \textit{strictly greater} than at least one neighbor are identified as source nodes (hubs); in the rare case, nodes having the same centrality score as all its neighbors are considered as isolated nodes \citep{Note1}. $S$ is the set of hubs, each of which is the core of an \textit{end-community}, and $|S|$ is the number of end-communities. 

Correspondingly, find the set of nodes whose centrality score is local trough, i.e. whose centrality score is smaller than all its neighbors. Among these nodes, those that are not leaf nodes (only having one neighbor) are identified as boundary nodes (sinks). Each node in this category has at least two neighbors that have great centrality scores who could pass the end-community label to it (see \textit{Step 2}), hence it belongs to more than one community; these nodes determine the boundary of end-communities. The remaining nodes in the graph are inner members who only belong to one community.

Therefore, based on the centrality measure relative to its neighbors, each node is identified with one of the five roles: hubs (sources), inner members, boundaries (sinks), isolated nodes and leaf nodes (Figure 1). Each hub defines an \textit{end-community}; hubs and inner members belong to a single community, while boundaries belong to multiple communities and serve as the overlaps between different communities. Isolated nodes are very rare and sparsely distributed in the graph and we assume that they don't have community membership; they could always be allocated to neighboring communities if one seeks to eliminate this category.

\textit{Step2}: \textbf{Determination of end-communities}. Assign a different community label $s$ on each hub, and initiate a diffusive label propagation process \textit{simultaneously} starting from all hubs in $S$. The membership of a specific end-community $s$ is represented by a tuple $x_s= \{(i,t)\}$, which records that node $i$ joins the community at time $t$. Correspondingly, every node $i$ is associated with an infection history (memory) tuple $h_i = \{(s,t)\}$ that records the label $s$ it receives at time $t$. The two tuples $X = \{x_s\}$ and $H= \{h_i\}$ are updated in the propagation process. Note that the synchronization of label propagation is guaranteed in our algorithm by using $t$ to record the timestamps of the label infection, instead of only recording the incident source of infection, as in \citep{G2010}.
 
To the first order, we assume that nodes only infect their immediate neighbors. The propagation rule is: at time $t$, starting from node $i$ with \textit{current} community label $s$, for a different node $j$, if $a_{ij} =1$ and $c_i > c_j$, add $(s,t)$ to $h_j$ and $(j,t)$ to $x_s$, when $(s, \forall t) \notin h_j$ (same as $(j, \forall t) \notin x_s$). In other words, there will be a successful infection of the community label, if and only if the incident node's centrality score is greater than the target node, an immediate neighbor of it, and the infection will be recorded when this is the first time the target node received this community label. The label propagation will not take place between two nodes having the same centrality score, which is consistent with our definition of isolated nodes (they are insulated from any infection).  

At each time step, the label propagation will spread to all neighbors of the newly infected nodes (except the infector of the previous step, since its centrality score is higher). The propagation of a certain label will stop at those directions where the neighbors to be infected have a higher centrality score, or the neighbors are already infected by that label (hence the infection history $H$ is non-repetitive). In the most extreme case where the graph has a strict tree structure of centrality scores, the label propagation will take at most $q$ time steps, where $q$ is the longest path length of the graph, and the length of the infection memory $h$ is at most $N-1$. In practice, the propagation could stop after only a few time steps, when all the community memberships $X = \{x_s\}$ (or equivalently, the infection history of all nodes $H = \{h_i\}$) do not change, i.e. no new community label is assigned to any node. The propagation process is sketched in Algorithm 1.

$H$ records the information of nodes' overlapping community membership. For node $i$, if $|H_i| > 1$, it is a boundary node that belongs to more than one community; if $|H_i| = 1$, node $i$ is either a source ($i \in S$) or an inner member ($i \notin S$).  From $X$ and $H$, we could qualify the overlappedness of communities in the graph by two metrics: (1) the average community membership of all nodes (average length of the non-repetitive infection history) $m_h$, and (2) the average size of end-communities $m_x$. The two metrics are related by:
\begin{equation}
m_h = \frac{|S|}{N} m_x.
\end{equation}

Notably, one advantage of recording the infection history $H$ is that, for nodes belonging to multiple communities, their strength to different communities could possibly be indicated by the infection time $t$: a small $t$ in $(s,t)$ means that the node is near to the hub $s$, thus having a large strength to this community, and vice versa.

\begin{figure*}
\begin{minipage}{\linewidth}
\begin{algorithm}[H]
  \caption{Propagatory formulation of end-communities (Step 2)}
  \label{alg1}
   \begin{algorithmic}[1]
  \State Initialize $X$, $H$.
  \State Input adjacency matrix $A= \{a_{ij}\}$ and calculate nodes' centrality scores $C = \{c_i\}$.
  \State Identify the set of hubs for end-communities $S$ (Step 1).
  \State $t = 0$
   \While{$t<t_{max}$}
   \State $t = t + 1$
	\For{$\forall$ community $s\in S$}
		\For{$\forall$ node $i$ s.t. $(i,t-1) \in x_s$}
			\For{$\forall$ node $j$ s.t. $a_{ij}=1$ and $c_i > c_j$}
				\If{$j$ not in community $s$}
					\State Add $(j,t)$ to $x_s$ 
					\State Add $(s,t)$ to $h_j$
				\EndIf
			\EndFor
		\EndFor	
	\EndFor
	\If{$H$ (or $X$) not change}
		\State \textbf{break}
	\EndIf
   \EndWhile
   \State $X$, $H$, $t_{fin}$ obtained. Calculate $m_h, m_x$.
   \end{algorithmic}
\end{algorithm}
\end{minipage}
\end{figure*}

\textit{Step3}: \textbf{Aggregation of small communities}. Communities join each other and form bigger ones when they are close enough. We assume that the distance between end-communities is represented by the distance of their hubs, and thus formulate the distance matrix $R_0$ over end-communities, where each entry $r^0_{pq}$ is the shortest path between two hubs $p$ and $q$:
\begin{equation}
R_0 = \{r^0_{pq} \} = \{d_{shortest\ path} (p,q) \},\ \ p,q \in S.
\end{equation}

If no path exists between $p$ and $q$ (in the case of sub-components; the graph is not fully connected), we set $r^0_{pq}= d_{inf}$ (in practice, $d_{inf}$ is an extremely large number). The distance matrix $R_0$ is $|S| \times |S|$, whose diagonal elements are all 0 and off-diagonal elements are positive integers. Now we aggregate end-communities by iteratively rearrange the distance matrix $R_0$. Find two hubs (two rows in the matrix) with distance $\epsilon = 1$ and replace them with a single merged node in the matrix (not on the graph); for any unaddressed node, take the maximum of the two original distances in the matrix $R_0$ as the new distance between the node and the aggregated new node. Repeat until all $\epsilon=1$ elements in the $R_0$ matrix are detected and replaced. This procedure iteratively reduces the size of $R_0$ and update the matrix; in the end, larger communities  ($\epsilon_1$-communities) are formed out of end-communities and we obtain the new distance matrix $R_1$. Using the same approach, we then formulate $\epsilon_2$-communities and $R_2$, up to the final $\epsilon_{d^{max}}$-community and $R_{d^{max}}=[0, d^{max};d^{max},0]$, where $d^{max}$ is the largest shortest path distance between the hubs in the graph, i.e. the largest element in $R_0$. For conveniences, we write $\epsilon_{d^{max}}$ as $\epsilon_{max}$ and $R_{d^{max}}$ as $R_{max}$. In the end, a hierarchy of communities is obtained through this upward iterative aggregation of small communities, whose distances are represented by a series of matrices $R_0$, $R_1$, ... $R_{max}$ of gradually reduced sizes. This step is summarized in Algorithm 2 and illustrated in Figure 3.

Note that our algorithm naturally takes care of input graphs that are not fully connected through $d_{inf}$: during the iterative reduction of $R$, once we find that at a certain stage, all off-diagonal elements of $R$ equal $d_{inf}$, it suggests that the remaining communities are the sub-components and could not be further combined, and thus the aggregation process stops, with $d_{max}$ indicating the largest diameter of the subcomponents of the unconnected graph.

For integer $\epsilon$ ranging from $\epsilon_0$ to $\epsilon_{max}$, a different number of communities (the size of $|R_{\epsilon}|$) remain at each value of $\epsilon$. The $\epsilon\leftrightarrow |R_{\epsilon}|$ relationship demonstrates the nature of the community hierarchy; turning points of the slope of the $\epsilon \leftrightarrow |R_{\epsilon}|$ curve could be used in practice to suggest the cutoff level of the obtained community hierarchy (see Results and Discussion).

\textit{Step4}: \textbf{Quality check of the hierarchy}. Calculate the Jaccard index matrix $J$ between each pair of end-communities whose hubs are $p,q$: $J = \{j_{pq}\}$, which is an index characterizing the similarity of two groups of nodes. The automatic quality check of the detection results is carried out relying on the Jaccard matrix $J$. In the previous step, we aggregate small communities based on the distance of their source nodes; conceptually, one may propose an alternative aggregation rule: iteratively aggregate small communities that have the most overlap, indicated by the Jaccard index. However, one problem arises with this plausible treatment: it may almost always lead to a strictly \textit{binary} hierarchy that at each step, only one big community will be formulated out of exactly two small communities, since the Jaccard index is a real number. By contrast, our (integer-valued) distance-based aggregation rule allows that at each stage a few mergings take place. In other words, an overlap-based merging rule will almost always result in a hierarchy with $|S| -1$ stages and therefore formulate a dendrogram, which is a binary-tree structure \citep{Cet2008}, whereas our distance-based merging rule is more flexible and may be able to yield a much tighter hierarchy (k-ary tree).

Nevertheless, we could utilize the Jaccard index to check the quality of our distance-based mergings. Under our distance-based rule, at each merging, i.e. combing two communities $p,q$ into one $p+q$, the Jaccard index $j_{pq}$ is not necessarily the largest element in the matrix $J$ (i.e. $p$ and $q$ do not necessarily have the most overlap among the pairing of all communities); however, to be considered a \textit{good} merging, one idea is that it should be satisfied that the Jaccard index between the two communities $p,q$ to be merged, must be larger than the index between one community out of $p,q$ and any other community at the current stage that is not going to be further merged with $p$ and $q$ (i.e. whose two distances to $p,q$ are not \textit{both} the same as $r_{pq}$). This means that, the merging of $p$ and $q$ will be considered \textit{good} (i.e., consistent with overlap-based heuristics), if and only if all the other communities that have more overlap with $p$ \textit{or} with $q$ are going to be further merged with $p$ and $q$ at this $\epsilon$ stage, or equivalently, no community that has more overlap is not to be merged. We call this condition as \textit{J-D consistency}, which is formally stated as:

\begin{widetext}
\begin{equation}
\text{J-D consistency}: j^{\epsilon}_{pq} > j^{\epsilon}_{pz} \ \ \text{and}\ \  j^{\epsilon}_{pq} > j^{\epsilon}_{qz}, \ \ \ \forall z\in HR_{\epsilon}\ \ s.t. \ \ r^{\epsilon}_{pz} > r^{\epsilon}_{pq} \ \text{or}\ r^{\epsilon}_{qz} > r^{\epsilon}_{pq} \ \ \text{(for a certain $\epsilon$)}.
\end{equation}
\end{widetext}

where $HR_{\epsilon}$ denotes the set of communities obtained at a certain $\epsilon$ level of the hierarchy (i.e., in the beginning, $HR_0 = S$; in the end, $|HR_{\epsilon_{max}}| = 1$ for connected graphs). If the above J-D consistency is satisfied, the merging at this step is considered as a good merging. Hence, by this means, we are able to indicate the quality of the obtained community hierarchy (thus the quality of our detection workflow) by a J-D consistency factor $\Phi$, which is the number of good mergings (condition (3) satisfied) normalized by the total number of mergings $|S|-1$. Since the last merging will always be J-D consistent, one number is subtracted from both the numerator and the denominator of the ratio, and we have:
\begin{equation}
\Phi = \frac{card(\text{J-D consistent})-1}{|S|-2}.
\end{equation}

Note that $\Phi$ is applicable only when $|S| > 2$, i.e. there are more than 2 end-communities detected in the beginning. If $\Phi = 1$, our distance-based merging rule is perfectly consistent with the overlap-based rule. During the merging process, the Jaccard index matrix $J$ is recalculated at each stage, and the dimension reduction of $J$ is in accordance with the dimension reduction of $R$ (Algorithm 2). In practice, each merging event is associated with a boolean variable indicating its J-D consistency, and therefore at each $\epsilon$, we are able to calculate the $\Phi_{\epsilon}$ at that level, which is a component of the final $\Phi$. The curve $\epsilon\leftrightarrow\Phi_{\epsilon}$ also helps the determination of the cutoff level of the community hierarchy (see Results). 

The metric $\Phi$ provides the algorithm with the desired property of self-falsifiability. A large $\Phi$ indicates that the establishment of the community hierarchy obtained from the detection workflow embodies a large proportion of good mergings, in the sense that two communities whose hubs are shortest-distanced are also having the largest overlap, so that the aggregation of them is double credited. By contrast, a small $\Phi$ implies that the aggregation process mostly consists of bad mergings, in which the two merging heuristics often do not coincide; therefore, our detection scheme may not be suitable for the specific graph.

\algdef{SE}[SUBALG]{Indent}{EndIndent}{}{\algorithmicend\ }%
\algtext*{Indent}
\algtext*{EndIndent}

\begin{figure*}
\begin{minipage}{\linewidth}
\begin{algorithm}[H]
  \caption{Determination of the community hierarchy (Step 3 $\&$ 4)}
  \label{alg2}
   \begin{algorithmic}[1]
  \State Calculate $R_0$, $J_0$. Obtain $\epsilon_{max}$ from $R_0$. 
  \State $\epsilon = 0$, $\Phi = 0$, $HR_0 = S$.
  \While{$\epsilon<\epsilon_{{max}}$}
  \State $\epsilon = \epsilon + 1$
  \While{true}
  \State Find the $\epsilon$-element of $R$, whose position is $(p,q)$
		\If{no $\epsilon$-element found}
			\State \textbf{break}
		\EndIf
  \State Update the community hierarchy $HR$:
		\Indent
   			\State Remove the communities indexed by $p$ and $q$ from $HR_{\epsilon-1}$.
			\State Merge the two communities into a larger community $p+q$ and add to the hierarchy.
		\EndIndent

  \State Update the distance matrix $R$:
		\Indent
			\State Remove the columns and rows $\{p,q\}$ from $R_{\epsilon-1}$.
			\State Add a new row and a new column denoting the merged community $p+q$.
			\State $d(r,p+q) = max(d(r,p) , d(r,q))$
		\EndIndent

		\If{J-D consistency satisfied}
			\State $\Phi = \Phi+1$
			\State Record the J-D consistent merging event.
		\EndIf

  \State Update the Jaccard matrix $J$.
		
  \EndWhile
  \State $R_{\epsilon-1} \rightarrow R_{\epsilon}$, $HR_{\epsilon-1} \rightarrow HR_{\epsilon}$, $J_{\epsilon-1} \rightarrow J_{\epsilon}$
  \State Calculate $\Phi_{\epsilon}$.
  \If{all off-diagonal elements of $R_{\epsilon}$ equal $d_{inf}$}
	\State \textbf{break}
  \EndIf

  \EndWhile
  
\end{algorithmic}
\end{algorithm}
\end{minipage}
\end{figure*}

\subsection*{Computational Complexity}

Consider the graph with $N$ nodes and $E$ edges. The time complexity for calculating the centrality measures is $O(N)$ for degree centrality and $O(N\text{log}(N))$ for eigenvector centrality. After the centrality scores of all nodes have been obtained, at Step 1, the identification of nodes' roles is realized by comparing each node's centrality score to all its neighbors; this procedure incurs a time cost $O(E)$. At Step 2 (Algorithm 1), during the last iteration, every node in every community is visited, with each visit accessing all the node's neighbors. This corresponds to $2E$ visits at this (last) iteration, and therefore the entire time complexity of Step 2 is $O(t_{fin}E)$. Step 3 $\&$ 4 are carried out at the same time in Algorithm 2. The time complexity of Algorithm 2 depends on the dimension of the matrix $R_0$, which is determined by the number of end-communities (i.e. number of hubs) $|S|$. The $|S| \times |S|$ matrix $R_0$ gradually degenerates into the $2\times2$ matrix $R_{max}$, with the minimum element in the $R$ matrix detected at each stage; therefore the time is upper bounded by $O(|S|^2)$, as is also the time complexity for calculating the shortest distance between the hubs of end-communities in the formulation of $R_0$.

After all, the computational complexity of the entire algorithm $t_{algo}$ is (using degree centrality at Step 1; assuming $|E| > N$):
\begin{widetext}
\begin{equation}
t_{algo} = O(N) + O(E) + O(t_{fin}E) + O(|S|^2) = O(t_{fin}E) + O(|S|^2).
\end{equation}
\end{widetext}

In practice (see Results), $t_{fin}$ is always very small, and one could often set up a small $t_{max}$ to let the detection finish early by cutting off nodes' membership to remote communities; this treatment will not influence the final detection results in most cases. $|S|$ is also very small, normally a tiny fraction of $N$, and $|S|^2$ is unlikely to exceed $N$. Therefore, the computational complexity of our algorithm is effectively $O(E)$ when using degree centrality as the measure (Figure S1) , which is fast on real networks where node connections are often not dense.

\subsection*{Computational Superiority}

A few advantages of our algorithm could be highlighted in computation. First, by recording the timestamps of the infections, the label propagation process in our framework is synchronized. This advantage prevents the numerical error incurred by unsynchronized algorithms (e.g. the original label propagation \citep{Ret2007}). Next, Step 2 and Step 3 of the algorithm could run in parallel after Step 1, although Step 4 and the determination of the cutoff level of communities still need to be carried out after Step 2 and 3 are finished. Last, as demonstrated, the computational time of our algorithm is linear with the number of edges, which is a desirable feature for its application on massive real-world social networks. This is achieved by (1) the one-way propagation (hubs to surroundings) of labels with stops at centrality sinks (Step 2), which is notably faster than existing label propagation algorithms without a one-way formulation (quasi-linear with the number of edges), and (2) the iterative dimension reduction of the distance matrix of communities, which in practice often takes only a few steps to degenerate into the final matrix.

\section*{Results}
We applied our community detection scheme to networks of a wide range of size and various sorts (Table 1). Degree centrality is used as the centrality measure in the detection; tests show that using eigenvector centrality often fails to identify enough hubs for end-communities, and the computational cost for calculating eigenvector centrality for large scale networks is often prohibitive. In our experiments, the propagation process (Step 2) converges after a few number of iterations ($t_{fin}<22$) on all tested networks. We used small real networks (Karate club network \citep{Z1977}, Dolphin network \citep{Luet2003}) and synthetic networks (LFR benchmark network \citep{LLet2008} and Erd\"os R\'enyi (ER) random network \citep{ER2011}) to demonstrate the detailed procedures of our detection scheme and show the important $\epsilon \leftrightarrow \Delta |R_{\epsilon}|$ and $\epsilon\leftrightarrow\Phi_{\epsilon}$ relationships constructed along the formulation of the community hierarchy (Figure 2-4). We then apply the algorithm to a panel of large real networks to carry out horizontal discriminative analysis. A number of notable features emerged from the detection results, which demonstrated the self-consistency and robustness of our algorithms; meanwhile, a few unexpected interesting phenomena regarding the intrinsic structure of networks are uncovered (Figure 5-6).

\begin{figure*}
\begin{minipage}{\linewidth}
\begin{table}[H]
\begin{ruledtabular}
\begin{tabular}{@{}|c|c|c|c|c|c|c|c|c|c|c|c|c|@{}}
\toprule
\textbf{Network}             & \textbf{$\#$Node} & \textbf{$\#$Edge} & \textbf{$\#$Hub} & \textbf{$\#$Boundary} & \textbf{$\#$Isolate} & \textbf{$\#$Leaf}& \textbf{$\#$Inner} & \textbf{$t_{fin}$} & \textbf{$m_h$} & \textbf{$m_x$} & \textbf{$\epsilon_{max}$} & \textbf{$\Phi$} \\ \midrule
\hline
\textbf{Karate Club}            & 34                & 78 & 2                    & 16                 & 0          & 1 & 15           & 5                   & 1.5           & 25.5       &2    & 1                            \\ \midrule
\textbf{Dolphin}           & 62                &  159 & 5                    & 12                 & 0              & 9 & 36             & 5                   & 2.08         & 25.8       & 4   & 1                            \\ \midrule
\textbf{LFR (3, 1.2, 0.1)} & 1000             &  2153 & 74                   & 324               & 6             & 5 & 591              & 8                   & 2.05         & 25.9       & 10  & 0.260                        \\ \midrule
\textbf{Facebook users}          & 4039          &   88234   & 5                    & 621                & 0         & 75 & 3338                  & 13                  & 2.00         & 1617.2        & 6  & 0.333                          \\ \midrule

\textbf{Enron email*}     & 36692           & 183831   & 483                 & 8640                 & 530         & 11211 & 15828                  & 16                   & 3.65         & 277.2        & 11  & 0.131                        \\\midrule

\textbf{Brightkite*}     & 58228           & 214078   & 682                 & 12259                 & 49         & 21157 & 24081                  & 12                  & 2.91         & 248.5        & 16  & 0.193                        \\\midrule

\textbf{CA-GrQc*}     & 5241           & 14484   & 298                 & 851                 & 185         & 1197 & 2710                  & 13                  & 3.74         & 65.8        & 13  & 0.128                        \\\midrule

\textbf{CA-HepTh*}     & 9875           & 25973   & 341                 & 2123                 & 184         & 2109 & 5118                  & 14                  & 8.00         & 231.6        & 15  & 0.124                        \\\midrule

\textbf{CA-HepPh*}     & 12006           & 118489   & 172                 & 2605                 & 170         & 1493 & 7566                  & 17                  & 1.59         & 110.9        & 11  & 0.159                        \\\midrule

\textbf{CA-AstroPh*}     & 18771           & 198050   & 185                 & 3909                 & 281         & 1282 & 13114                  & 8                  & 1.94         & 196.8        & 11  & 0.082                        \\\midrule

\textbf{CA-CondMat*}     & 23133           & 93439   & 442                 & 4717                 & 447         & 2373 & 15154                  & 15                  & 9.38        & 490.9        & 13  & 0.107                        \\\midrule

\textbf{Deezer-RO}     & 41773           & 125826   & 1051                 & 9221                & 5         & 5430 & 26066                  & 18                  & 31.10       & 1236.3        & 17  & 0.119                        \\\midrule

\textbf{Deezer-HU}     & 47538           & 222887   & 450                 & 10494                & 0         & 2701 & 33893                 & 20                  & 112.31      & 11864.7        & 12  & 0.054                        \\\midrule

\textbf{Deezer-HR}     & 54573           & 498202   & 64                 & 11035                & 1         & 2330 & 41143                 & 19                  & 40.43      & 34473.1        & 10  & 0.145                        \\\midrule

\textbf{FB-artist}     & 50515           & 819090   & 30                 & 14570                & 0         & 3124 & 32791                 & 11                  & 3.96      & 6673.7       & 10  & 0.214                        \\\midrule

\textbf{FB-new sites}     & 27917          & 205964   & 179                 & 7762                & 0         & 2137 & 17839                 & 18                  & 14.33      & 2234.4       & 12  & 0.254                       \\\midrule

\textbf{FB-company}     & 14113          & 52126   & 341                 & 3602                & 3         & 2358 & 7809                 & 18                  & 25.38      & 1050.4       & 13  & 0.201                        \\\midrule

\textbf{FB-athletes}     & 13866          & 86811   & 43                 & 4715                & 0         & 1240 & 7868                 & 19                  & 16.72      & 5391.7       & 8  & 0.171                        \\\midrule

\textbf{FB-government}     & 7057          & 89429   & 15                 & 1894                & 0         & 355 & 4793                 & 15                  & 4.15      & 1951.8       & 9  & 0.385                        \\\midrule

\textbf{FB-politician}     & 5908          & 41706   & 60                 & 1845                & 0         & 600 & 3403                 & 14                  & 8.03      & 790.3       & 12  & 0.155                        \\\midrule

\textbf{FB-public figure}     & 11565          & 67038   & 129                 & 3239                & 0         & 1912 & 6285                 & 16                  & 6.83      & 612.7       & 13  & 0.268                        \\\midrule

\textbf{FB-tv show}     & 3892          & 17239   & 153                 & 997                & 0         & 611 & 2131                 & 11                  & 3.62      & 92.2       & 17  & 0.291                        \\\midrule

\textbf{Gowalla}     & 196591           & 950327   & 1266                 & 49295                 & 9         & 49452 & 96569                  & 22                  & 3.97         & 616.5        & 14  & 0.157                        \\\midrule

\textbf{Amazon}     & 334863           & 925872   & 17837                 & 120277                 & 71         & 25709 & 170969                  & 14                   & 2.30         & 43.2        & 16  & 0.193                        \\\midrule

\textbf{DBLP}     & 317080             & 1049866   & 2965                  & 68403                 & 1         & 43181 & 202530                  & 25                   & 56.1         & 5999.2      & 19   & 0.156             \\ \midrule

\textbf{ER (p = 0.1)}     & 50             & 144 & 4                   & 10                 & 0                & 0 & 36           & 5                   & 2.64         & 33.0         & 2  & 0.5                          \\ \midrule

\textbf{ER (p = 0.01)}     & 498             & 1279   & 44                   & 112                 & 0         & 23 & 319                  & 8                   & 4.07         & 46.1        & 5  & 0.095                        \\ \midrule

\textbf{ER (p = 0.001)}     & 4961             & 12307 & 491                   & 1161                 & 0                & 166 & 3143           & 9                   & 4.59         & 46.4         & 9  & 0.065                          \\ \midrule

\textbf{ER (p = 0.0001)}     & 49682             & 124911 & 4692                   & 11615                 & 1                & 1725 & 31649           & 11                   & 4.87         & 51.6         & 13  & 0.047                          \\ \bottomrule

\end{tabular}%
\end{ruledtabular}
\caption{Summary of detection results. Networks with star marks are not fully connected. On each network, self-edges and nodes with degree 0 are removed, a trivial modification to the original graph in all cases.}
\end{table}

\end{minipage}
\end{figure*}

\textbf{\textit{Karate club network}}

The two centers (node $\#$0, Mr. Hi; node $\#$33, the officer) in the network are successfully identified as the only two hubs (blue nodes; Figure 2, left top), around which two end-communities (obviously, the only non-trivial communities in the 2-level hierarchy) are determined (Figure 2, left bottom). Although our algorithm detects overlapping communities while the ground truth communities of the Karate club network are disjoint, the detection results recover the ground truth to a great extent (Figure 2, right). First, the two overlapping communities of our results (second and fourth column) strictly contain the ground truth (first and third column). Second, as in our detection process each community assignment is associated with a timestamp, one may decide that the earlier the node joins the community, the larger its strength to this community is. Therefore, by abandoning the nodes that have lower strength to the communities (i.e. nodes having large values in the timestamp), it is possible to further compare the (truncated) overlapping communities with the ground truth. Specifically, when only considering the nodes that join the community before time $t=2$, our detected communities deviate from the ground truth by a small margin (entries in red; Figure 2, right).

\begin{figure*}
\begin{minipage}{\linewidth}
\centering   
	\includegraphics[width=6in]{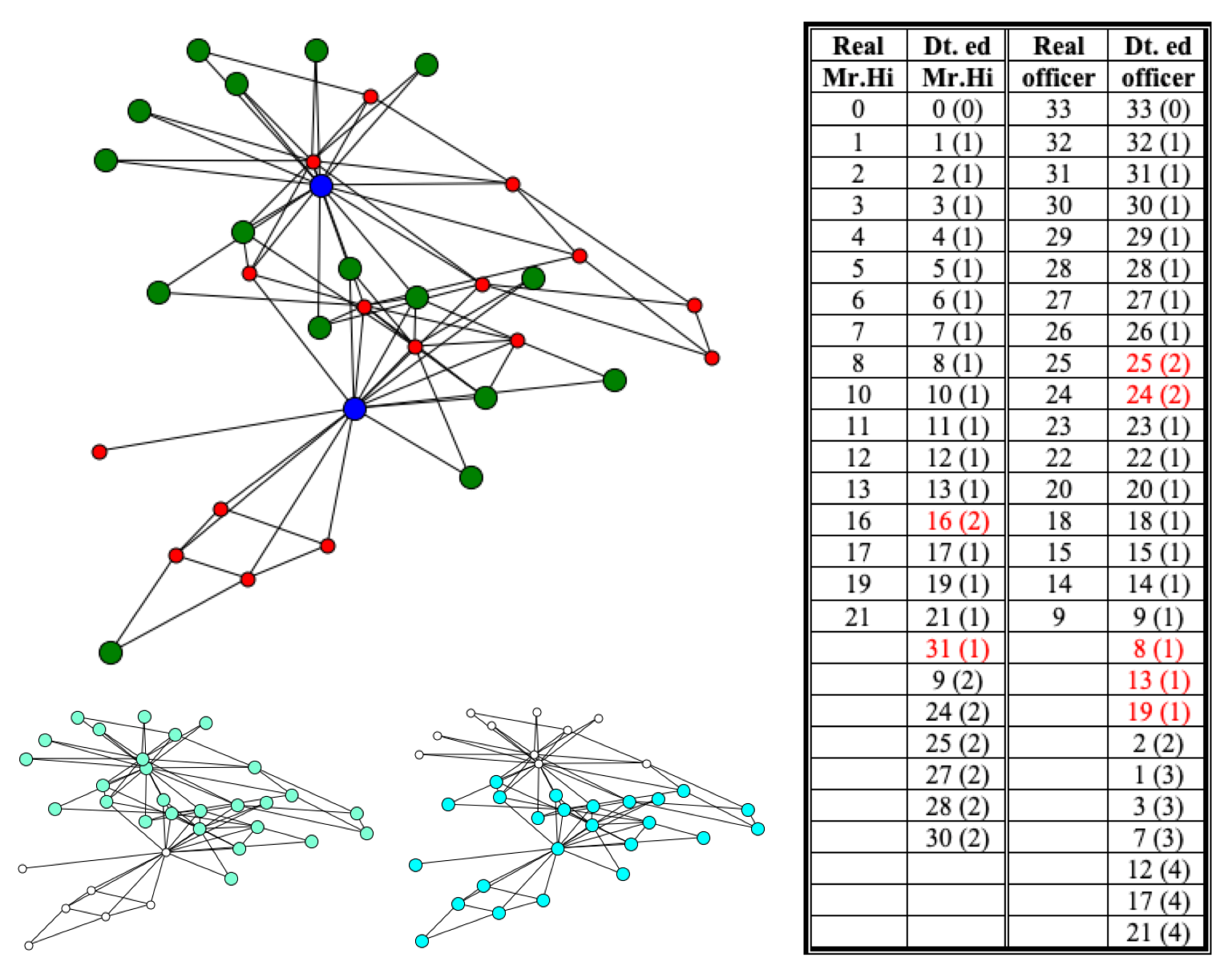}  
\caption{Detection results on the Karate club network \citep{Z1977}. Left top: nodes' different roles. Blue, red, green nodes are hubs, inner members and boundaries, respectively. Left bottom: two end-communities detected. Right: detection results compared with the ground truth.} 

\end{minipage}
\end{figure*}

\textbf{\textit{Dolphin network}}

The Dolphin network contains $62$ nodes, among which $5$ are identified as hubs (blue nodes; Figure 3, middle top) for the corresponding 5 end-communities (Figure 3, leftmost panel). The iterative reduction of the $5 \times 5$ distance matrix $R_0$ and the sequential aggregation of small communities are demonstrated in detail (Figure 3, rightmost panel). Red marks show the communities that are aggregated at each stage of $\epsilon$. The community hierarchy is obtained at the end of this iterative process (Figure 3, middle bottom). In each of the four merging event, the J-D consistency condition is satisfied (right marks; Figure 3, middle bottom), and thus the consistency factor $\Phi=1$. From the $\epsilon \leftrightarrow \Delta |R_{\epsilon}|$ relationship, one can see that a proper cutoff level for communities is $\epsilon=2$, and the corresponding two $\epsilon_2$-communities are shown. Such a cutoff is chosen because the community membership does not change at the following $\epsilon=3$ level, indicating that $\epsilon=2$ may be a characteristic distance between communities. This example shows that local peaks on the $\epsilon \leftrightarrow \Delta |R_{\epsilon}|$ curve could be considered as the cutoff level on the final community hierarchy. 

\begin{figure*}
\begin{minipage}{\linewidth}
\centering   
	\includegraphics[width=6in]{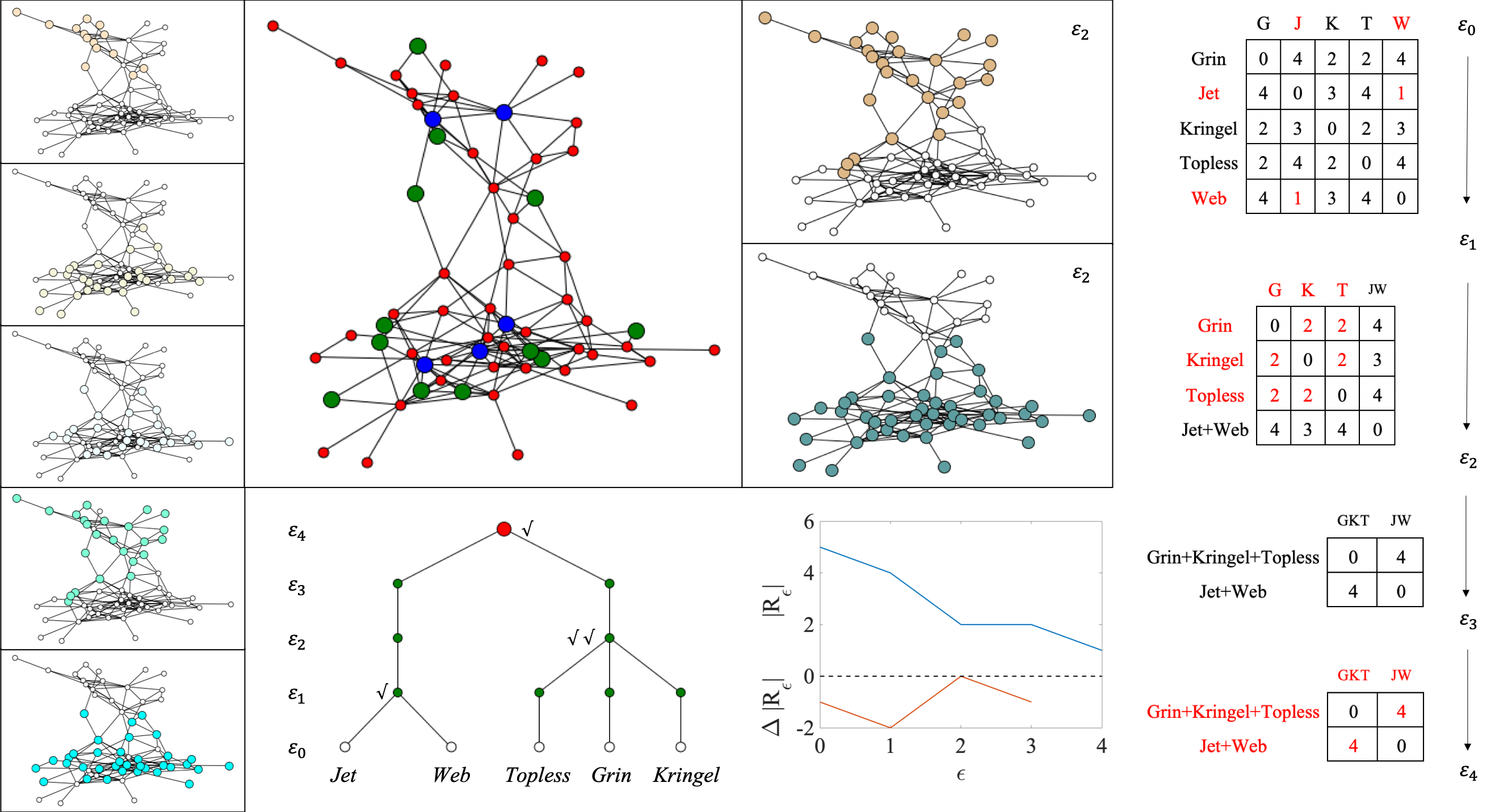}  
\caption{Detection results on the Dolphin network \citep{Luet2003}. Leftmost: 5 end-communities. Rightmost: iterative reduction of the distance matrix $R$ and aggregation of small communities. Middle top: nodes' different roles and the two $\epsilon_2$-communities. Blue, red, green nodes are hubs, inner members and boundaries, respectively. Middle bottom: the obtained community hierarchy and the $\epsilon \leftrightarrow |R_{\epsilon}|$ relationship.} 
\end{minipage}
\end{figure*}

\textbf{\textit{LFR benchmark network}}

We applied our detection algorithm to a LFR benchmark network of 1000 nodes ($\tau_1 = 3,\tau_2 = 1.2,\mu=0.1$) with 17 synthetic communities. Our algorithm identified $74$ end-communities (blue nodes; Figure $4$, left) during a propagation process of $8$ time steps. The formulated complete community hierarchy demonstrates the gradual build-up of large communities from smaller ones (Figure 4, right top). The $\epsilon \leftrightarrow \Delta |R_{\epsilon}|$ and $\epsilon\leftrightarrow\Phi_{\epsilon}$ relationships are obtained along the formulation of the hierarchy (Figure $4$, right bottom). It showed that from $\epsilon = 5$ to $\epsilon = 4$, the hierarchy experienced the least change (red curve), and the J-D consistency factor also arrived at a local peak during the merging from $\epsilon = 6$ to $\epsilon = 5$ (yellow curve). They both suggest that $\epsilon = 5$ is a good cutoff level, at which stage there are $16$ communities present; this is a good recovery of the synthetic ground truth ($17$ communities). A number of other LFR networks were synthesized and tested; in general our detection algorithm yielded similar performances.

\begin{figure*}
\begin{minipage}{\linewidth}
\centering   
	\includegraphics[width=6in]{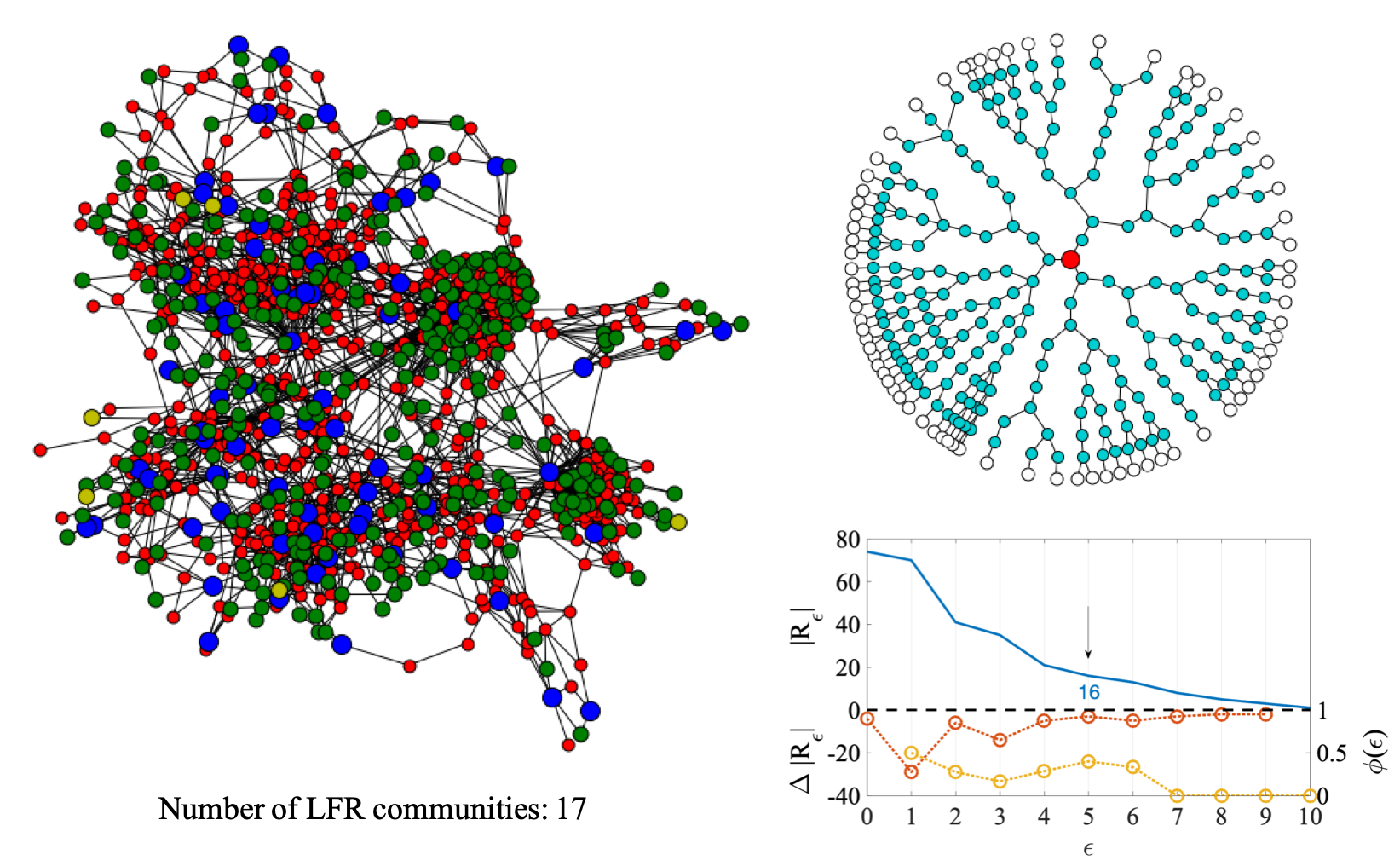}  
\caption{Detection results on the LFR benchmark network ($N = 1000, \tau_1 = 3,\tau_2 = 1.2,\mu=0.1$) \citep{LLet2008}. Left: nodes' different roles. Blue, red, green, yellow nodes are hubs, inner members, boundaries and isolated nodes, respectively. Right top: the obtained community hierarchy, built up from 74 detected end-communities (white nodes) to the entire graph (the red node). Lower right: $\epsilon \leftrightarrow \Delta |R_{\epsilon}|$ (red) and $\epsilon\leftrightarrow\Phi_{\epsilon}$ (yellow) relationships. An optimal cutoff level could be determined at $\epsilon = 5$.} 
\end{minipage}
\end{figure*}

\textbf{\textit{Erd\"os R\'enyi (ER) random network}}

We tested our detection algorithm on ER networks; a reasonable community detection method should be able to discover that these network do not contain significant community structures. Multiple ER networks are synthesized, with $(n,p)$ selected such that the number of nodes and edges of the synthetic networks are close to the magnitude of the real networks in use, in order to make fair comparisons (Table 1). Results show that, as desired, our detection scheme clearly separates random networks from real networks, which presumably have certain community structures embedded (Figure 5). First, repeated tests show that the proportion of hubs identified among all nodes (i.e., $|S|/N$) is always significantly smaller for real networks than random networks of similar sizes (Figure 5b), as one would expect, since random networks have a relatively flat structure and thus many nodes would be identified as ``plain hubs''. No similar distinction emerges in the proportion of boundary nodes and inner members, where random networks and real networks are indistinguishable from each other (Figure 5c, 5d). Second, for random networks the J-D consistency condition is poorly matched; $\Phi$ is small compared with real networks of similar size (Figure 5a). This suggests that, unsurprisingly, on random networks, not only is the identification of end-communities (hubs) unwarranted, but also the merging of these end-communities not self-consistent. One might also be able to spot random networks during the propagation process in Step 2; real networks typically show an S-shape in the cumulative iteration time plot, while random networks have a flatter running time growth (Figure S1). The two quantities $|S|/N$ and $\Phi$ could thus be used as the self-falsifiability benchmarks for detection results: for an arbitrary network, issue a random network of similar size and carry out detection on the two networks; if either $|S|/N$ or $\Phi$ in the detection result of the original network falls below the value of that on the issued random network, one should realize that the detection is not valid and the algorithm should be considered as not suitable for this specific graph.

\textbf{\textit{Large real networks}}

We also tested our algorithm on a number of large real networks across a wide range of magnitude, including the DBLP network and Amazon product network \citep{YL2015}, the Enron email network \citep{Leet2009}, the Facebook user network \citep{LM2012}, the five Arxiv collaboration networks \citep{Leet2007}, the recent data from two digital platforms (Deezer, $3$ networks; Facebook, $8$ networks) \citep{Ret2018}, and the Gowalla network and the Brightkite network, both location-based social networks \citep{Cet2011}. Detection results are summarized in Table 1, and a few critical metrics are visualized in Figure 5 and 6. The cutoff level of the community hierarchy for large real networks could be determined from the $\epsilon \leftrightarrow \Delta |R_{\epsilon}|$ and $\epsilon\leftrightarrow\Phi_{\epsilon}$ relationships, in the same way as for small networks (Figure S2). Yet It is difficult to make further discussion on the hierarchy cutoffs based on the current information; hence we focus on the horizontal discriminative analysis of the detection results on various networks.

One very interesting result is that the proportion of boundary nodes, identified in Step 1, exhibits a very small variance across all networks that we have studied, with an average value $25.0\%\pm6.5\%$ (Figure 5c). In our definition, boundary nodes are those having multiple community memberships; therefore, this result may imply that, on average, around 1/4 of nodes belong to multiple communities (i.e., they are ``crossovers'') on many kinds of real networks. No similar phenomenon could be seen in the proportion of either hubs (Figure 5b) or inner members (Figure 5d). Moreover, although this result is surprisingly robust across various real networks, tests show that it does not always hold true (as one would expect) for ER random networks of different $(n,p)$ and may depend on their $|E|/|N|$ values. At the current stage, however, no structural explanation could be warranted for this observation, and further analysis need to be carried out to better understand this phenomenon.

The categorical data facilitate the comparison of our algorithm's performance on (digital) social networks (3 Deezer networks, 8 Facebook networks) and on traditional (communication) networks (5 ArXiv collaboration networks). Results show that (Figure 6), the average size of community membership ($m_h$) and the average size of each community ($m_x/N$, as a proportion of the network size) of social networks (green and orange) are both clearly greater than that of traditional networks (grey). This is consistent with empirical considerations: on digital social networks, nodes have more access to different communities and thus it is easier to join multiple groups online than offline. Comparisons between different facebook groups are further indicative (Figure 6a): the average size of community membership is significantly smaller on artist, government and tv-show networks than on politician, athlete, company and public-figure networks, which is close to what one would imagine in real-world situations. As mentioned, it is expected that our algorithm will be more suitable for social networks than traditional networks, i.e. the quality factor $\Phi$ on social networks will be larger; unfortunately, while results clearly do not show the other way around, the winning margin is relatively vague (Table 1 and Figure 5a). Last, for Step 2 and 3 of the detection scheme, results show that both the propagation time $t_{fin}$ and the largest distance between end-communities $\epsilon_{max}$ (and the running time as well, Figure S1) are in general positively correlated with the network size (Figure 6c), which is consistent with our expectations.

\begin{figure*}
\begin{minipage}{\linewidth}

\centering   
	\includegraphics[width=6.5in]{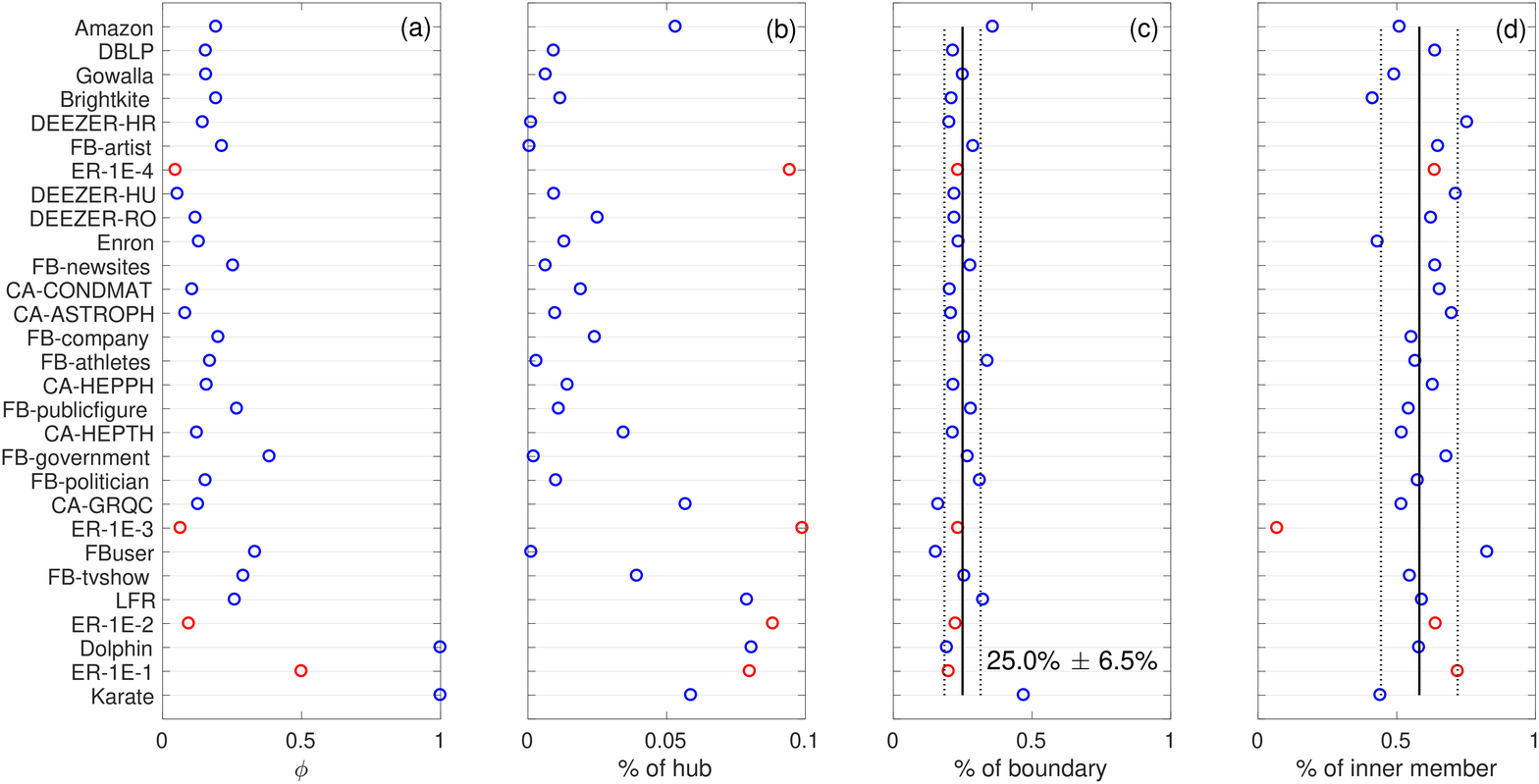}  
\caption{Identification of nodes' different roles (Step 1) and the J-D consistency factor (Step 4) on various networks. Random networks shown in red; real networks in blue. (a) $\Phi$, (b) proportion of hubs ($|S|/N$), (c) proportion of boundary nodes, (d) proportion of inner members. $\Phi$ and $|S|/N$ can effectively signify random networks. Interestingly, the proportion of boundary nodes in the graph shows a very small variance across all networks and centers around 25$\%$. No similar feature exists in the proportion of either hubs or inner members.} 
\end{minipage}
\end{figure*}

\begin{figure*}
\begin{minipage}{\linewidth}

\centering   
	\includegraphics[width=6.5in]{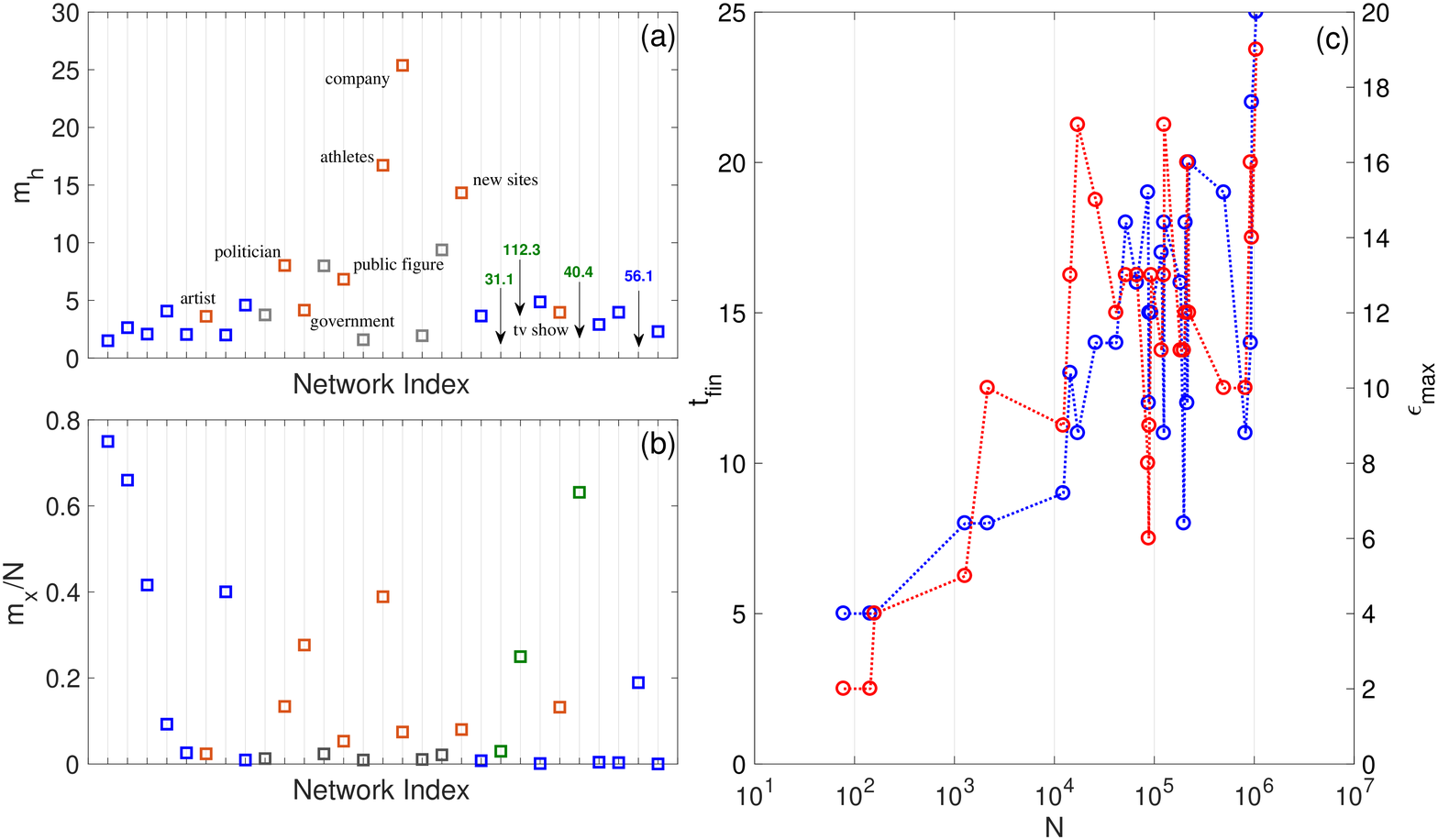}  
\caption{Critical metrics of the obtained community hierarchy. (a) Average community membership of each node; (b) average community size, as a proportion of the size of the network. Green: Deezer networks; orange: Facebook networks; grey: ArXiv collaboration networks. Names of the Facebook network category are labeled in text. (c) Propagation timestep $t_{fin}$ (blue, left axis) and the largest distance between end-communities $\epsilon_{max}$ (red, right axis), as a function of the network size (in logarithm).} 
\end{minipage}
\end{figure*}

\textbf{\textit{Comparison with other detection methods}}

We compare our detection results with the results of a few well-known algorithms for overlapping community detection, including the clique percolation (Perco) method \citep{Pet2005, Ret2012}, the link community (HLC) method \citep{Aet2010}, the SLPA algorithm \citep{Xet2011}, and the DEMON algorithm \citep{Cet2012}; both SLPA and DEMON adopt the label propagation process, which our detection scheme relies on as well. Recommended parameters are used for these reference algorithms: for SLPA, the iteration timestep is 20 and $r=0.1$; for DEMON, $\epsilon_{DEMON} =0.25$ and the minimum community size is 3; for Perco, $k=4$ (4-clique); for HLC, the dendrogram is not cut and the threshold is not used. Facebook networks (8 networks) and Deezer networks (3 networks) are used to carry out the comparison; these two groups of networks are from the same data source. A random network ER(5e3,1e-3) is also initiated for the experiments.

The performance of different algorithms are shown and compared in Figure 7. Generally speaking, our algorithm detects fewer but larger communities: among all, its results contain the smallest number of communities with the largest average community size $m_x$. Note that here we plot the number of end-communities (hubs) in our detected hierarchy; high-level communities ($\epsilon>0$) are  even larger and more scarce. This suggests that our detection scheme identifies much denser community structures on networks than the other four algorithms. The small variance across various networks on the proportion of boundaries among all nodes, which is mentioned earlier, is a unique feature of our algorithm; results of other algorithms show large variances on this metric. 

Tests suggest a few advantages of our detection scheme. Perco and DEMON could not process properly on the sparse random ER network, and HLC did not generate result on the large size network (facebook-artist) even after a long computational time; corresponding detection results are missing in Figure 7. By contrast, our algorithm is robust on both sparse and large-scale networks. SLPA is not deterministic, and detection results from multiple runs differ to a non-trivial extent; it is also not able to clearly separate random networks from real networks, at least by the number of communities detected as a fraction of the number of nodes ($\%$ of hubs), which is considered as an important metric in the detection results of our algorithm (Figure 7b). Perco often assigns no community to a large portion of nodes, given the sparse existence of cliques in real networks; even so, it found more communities than our algorithm, most of which are small-scale. Given its special nature, the link community method (HLC) always discovered more communities than the number of nodes in the network (i.e., $|S|/E<1$, but $|S|/N>1$) and thus corresponding results are omitted in Figure 7b; various tests suggest that HLC is not reliable for the assignment of community membership on nodes, as opposed to on edges. Both Perco and HLC determined communities far smaller than our algorithm; they also do not exhibit consistent $O(E)$ time complexity, unlike SLPA and DEMON (Figure S2). In a further test, we examined the performance of these algorithms on the Karate club network, which clearly shows that our detection results are the most reliable on this classic small network (Figure S4, Appendix D). In general, DEMON yields detection results closest to the results of our algorithm, yet its inability on sparse networks (Figure 7) and relatively insufficient coverage of nodes in communities (e.g. Figure S4) highlight the advantage of our new detection scheme. Finally, none of these reference algorithms has the ability to self-indicate its effectiveness on different networks and they all rely on certain parameter-tuning efforts in practice. The two novel features (parameter-free and self-falsifiable) of our solution scheme stand out.

\begin{figure*}
\begin{minipage}{\linewidth}
\centering   
	\includegraphics[width=6in]{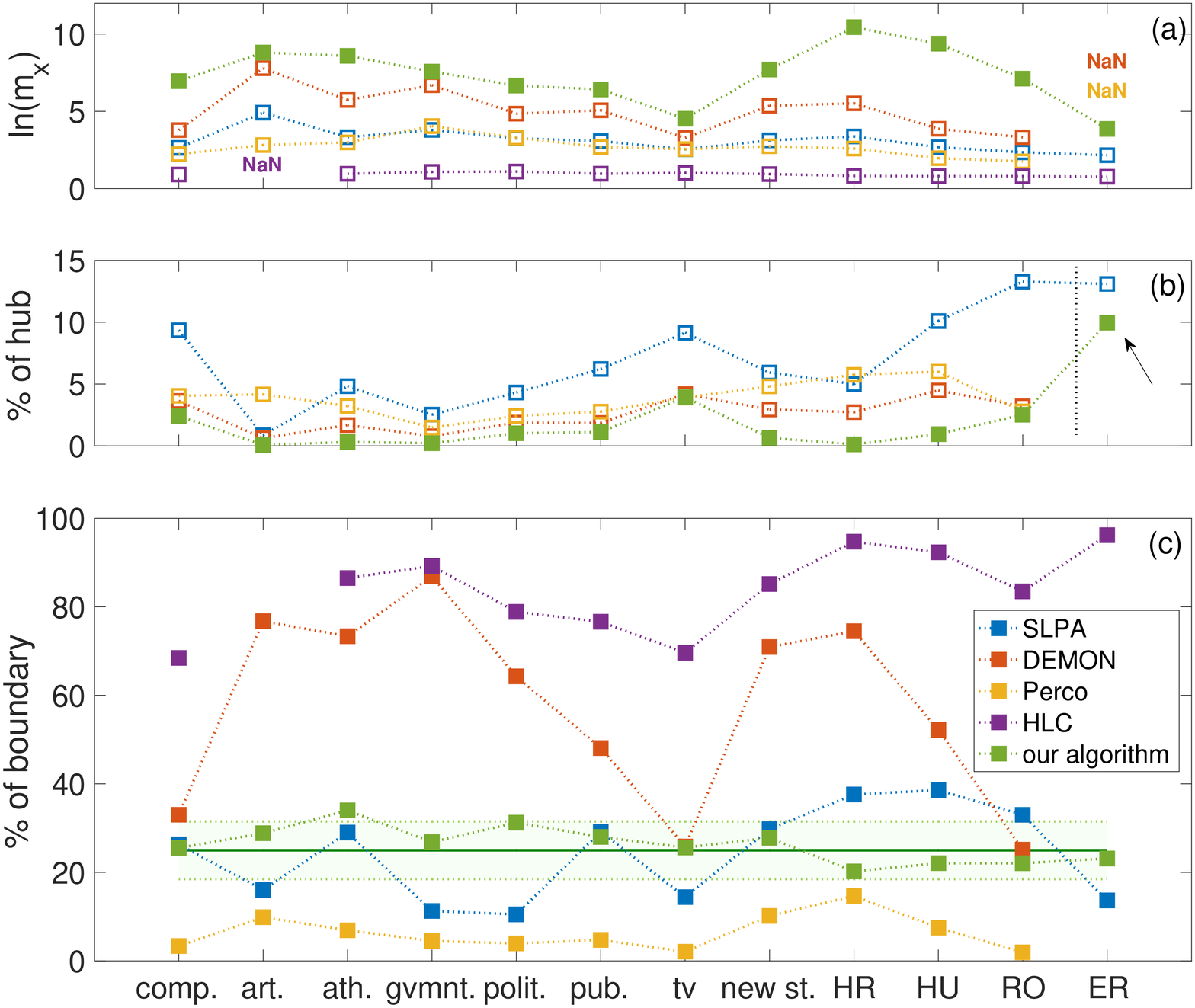}  
\caption{Comparison with other detection methods. 8 Facebook networks and 3 Deezer networks plus a sparse random network ER(5e3,1e-3) are used to demonstrate the performance of different algorithms, showing (a) $ln(m_x)$, (b) proportion of hubs (i.e., number of detected communities normalized by network size), and (c) proportion of boundary nodes (i.e., proportion of nodes that have multiple community memberships). Four well-known algorithms are studied besides our algorithm: clique percolation (Perco) \citep{Pet2005, Ret2012}, link community (HLC) \citep{Aet2010}, SLPA \citep{Xet2011}, and DEMON \citep{Cet2012}. Perco and DEMON could not process properly on the sparse random network, and HLC did not generate result on the large-scale network (FB-artist); corresponding detection results are missing (NaN). Given its special nature, link community (HLC) discovered more communities than the number of nodes in the network (i.e., $|S|/E<1$, but $|S|/N>1$) and thus the results are omitted in (b). Our algorithm successfully separates the random network from real networks, a feature not retained by SLPA (black arrow in (b)). In general, our algorithm detected fewer but larger communities in the graph. The small variance across various networks on the proportion of boundaries among all nodes is a unique feature of our algorithm (green shaded area; $25.0\%\pm6.5\%$).} 
\end{minipage}
\end{figure*}

\section*{Discussion and Concluding Remarks} 

In this study, we formulated an integrated belief for the algorithmic design of the community detection problem, consisting of six aspects: \textit{overlappedness}, \textit{different roles of nodes}, \textit{behavioral locality}, \textit{propagatory formulation of communities}, \textit{order of communities}, and \textit{self-falsifiability}. Based on the belief, we proposed a multi-step detection scheme that tries to incorporate successful ideas of existing algorithms as well as to obviate their exposed weaknesses. Our solution scheme relies on nodes' centrality scores to determine their different roles in the graph, especially the hubs and boundaries of end-communities, and initiates a diffusive label propagation process that tries to simulate the formation of communities on social networks. Small communities are iteratively aggregated into large communities and at the end of the detection, a hierarchical order of overlapping communities is established, with the entire graph sitting on the top of the hierarchy. Since there is in fact no concrete and general definition of a community structure on graphs and communities could then only be defined in the relative sense \citep{Pet2017}, we are attached to the belief that the old problem of finding the best partition of communities could be replaced by the new problem of finding the best cutoff level on a community hierarchy, which could be constructed on any given graph. With this idea in mind, in this study our solution scheme makes a tentative attempt. 

Our detection algorithm is parameter-free, and therefore as a trade-off, it is not fully decisive. While consolidated detection results of community structures are not produced by our completely objective algorithm, we adopt a few sophisticated measures that provide useful information for the determination of communities, specifically, the cutoff level of the community hierarchy. A peak on the $\epsilon \leftrightarrow \Delta |R_{\epsilon}|$ curve (or equivalently, a plateau on the $\epsilon \leftrightarrow |R_{\epsilon}|$ curve) means that across a certain $\epsilon$ stage the community hierarchy barely changes, which implies that such an $\epsilon$ level might be an appropriate candidate for the cutoff. Similarly, a peak on the $\epsilon\leftrightarrow\Phi_{\epsilon}$ curve suggests that across such $\epsilon$ level the merging of small communities into big ones is well-conditioned, in terms of the defined J-D consistency (equation (2)); thus this $\epsilon$ level is also a desired cutoff. By taking into account these two aspects, which often agree on the same $\epsilon$, we may be able to decide an appropriate cutoff level of the community hierarchy. However, it should be noted that despite the proposed solution, the determination of the cutoff level is far from being consolidated; in many cases, subjective heuristics still need to be called for in making the decision.

An important feature of our detection scheme is the automatic indication of the goodness of detection results. As discussed in \citet{Pet2017}, any community detection algorithm has only a limited power in application, inevitably not being able to conduct successful detections on networks with certain topologies. Therefore, we believe that a reliable detection scheme should be able to notify implementers with the quality of the detection results it yields; in particular, the scheme should be able to indicate its potential failures. Such an automatic self-check procedure is embedded in our algorithm. By defining the concept of J-D consistency which indicates the quality of the mergings of small communities into big ones during the formation of the community hierarchy, we invented a robust metric $\Phi$ that quantitatively indicates the quality of detection results (which may also facilitate the determination of the best cutoff level on the community hierarchy). Self-falsifiability and the parameter-free property are not emphasized by existing algorithms and may be considered as novel features of our detection scheme.

We tested our algorithm on networks of various sizes and kinds (Figure 2-6). On small real networks (e.g. Karate Club network, Dolphin network), our algorithm yielded very good detection results. On LFR networks, our result is close to the ground truth, and it shows that our heuristics for determining the cutoff level of the community hierarchy are reliable. ER random networks could be effectively distinguished by our algorithm. On large-scale real networks, horizontal analysis further demonstrate the self-consistency of our detection scheme, and a few interesting phenomena emerge from the results, which exhibits extra values of this study beyond the algorithmic design. Specifically, an unexpected observation emerged, showing that under our identification scheme there are always around 1/4 nodes in the graph that belong to multiple communities, on various types of real networks. Although this phenomenon is significant in our results, more work needs to be done before it could be verified and generalize.

Advantages of our algorithms over existing overlapping community detection algorithms could be identified (Figure 7). Unlike the clique percolation method, the DEMON algorithm and the link community method, our detection scheme is robust on both sparse networks and large-scale networks; it yields deterministic detection results and successfully separates random networks from real networks, two superior features over the SLPA algorithm. In general, our algorithm generates fewer but larger communities than all the above algorithms, capturing the dense community structures on the network. The comparison of different algorithms' performance on the Karate club network provides unambiguous evidence in favor of our algorithm's reliability.


In our detection results, the strength of nodes' membership in different communities is not assumed to be homogenous and could possibly be indicated by utilizing the timestamp $t$ in their infection history, which records the first time the node gets exposed to community labels. The hierarchical order of communities are maintained throughout the workflow, thus the whole detection process is fully transparent. We believe that transparency is an important feature of this detection scheme, and the inclusion of timestamps in the finite memory associated with each node makes the algorithm easy to be extended to temporal networks or high-order networks \citep[e.g.][]{Lamet2018}, possibly with a refined centrality measure for these advanced networks \citep{Get2011}. Another line of extension for this study is to replace some flexible components of the algorithm and test with alternatives, for example, different centrality measures (Step 1) and alternative graph distance measures (Step 3). In the current scheme we used the most common measures (degree centrality, shortest path distance), but under the rapid development of network sciences, it would be interesting to apply and test alternative ideas under our general solution scheme in future studies.

A number of limitations exist in this study, besides what have been discussed. First, in theory, our algorithm is not able to identify communities that are \textit{strictly contained} in larger communities. Spectral clustering on the connectivity matrix of each determined community needs to be performed in order to find sub-communities strictly lying within big communities. Second, although we claim that the algorithm is parameter-free, a few quantitative constraints are still implied in our solution scheme, although they are not represented by explicit parameters. For example, we assume that nodes could only directly propagate the labels to their \textit{immediate} neighbors; this could be viewed as a dummy parameter $d_{prop}=1$ (distance of infections). The choice of centrality measures may also be viewed as a tuning procedure. Third, besides comparing on some general metrics of the detected communities, we found it a bit difficult to compare our detection results (a hierarchy of communities) with results obtained from other algorithms (a certain community partition) or more importantly, with the ground truth, although people argue that comparing detection results with ground truths may not be always desirable since the ground truth does not always reflect the real community structures of the network \citep{Pet2016}. It is plausible that we could compare the determined communities at the cutoff level of the community hierarchy with the singular detection result of other algorithms or the ground truth (e.g. by using the NMI index), but it is possible that multiple cutoffs could be identified in the hierarchy and therefore the comparison becomes less straightforward. Sophisticated metrics need to be invented to address this comparison, or in general, to better characterize the performance of our proposed solution scheme.

\subsection*{Data and Code Availability}
All network datasets used in this study could be found on Networkx (https://networkx.github.io) and SNAP (Stanford Network Analysis Project; http://snap.stanford.edu/index.html). A Python package of the detection algorithm is available.



\section*{Appendix A: Overview of Overlapping Community Detection Methods}

The topic of community detection on graphs is extensively studied over the time and a numerous set of algorithms have been proposed to deal with the problem. Although the research history for community detection is not long, there has seen multiple generations of views and ideas for this topic, and traditional methods are quickly surpassed by more advanced approaches. Important transitions of ideas include the transition from detecting exhaustive (disjoint) communities to overlapping communities, the transition from a deterministic definition of communities to a probabilistic definition of communities, and the transition from relying on synthetic data and data of small real networks without explicit community structures to test the algorithm, to utilizing networks with ground truth community structures, and then to further realizing the limits of ground truth constraints in evaluating community detection results \citep{FH2016}. The active evolution of community detection methods reflects the unconsolidated nature of the problem. 

As discussed in the main text, existing algorithms offer a great number of important aspects for the algorithmic design on overlapping community detection. From an evolutionary perspective, those ideas constitute a transitional logic line for thinking about the problem, and one could identify multiple stages in the development of solutions. Here we present an overview of overlapping community detection methods, trying to establish conceptual links connecting different ideas and to point out their successful insights as well as shortcomings.

\subsection*{Link Communities}

The idea of link communities, detected by a hierarchical clustering of edges \citep{Aet2010}, is based on the assumption that vertex communities may be overlapped but the corresponding link communities are always disjoint. In other words, it implies that the boundaries of communities are not determined by nodes, as traditionally assumed, but by the edges connecting them. Despite being an advanced view over hard-partitioning of nodes, this idea is still subject to improvements since it is possible that edges also belong to different communities and hard-partitioning on edges is still an imposed assumption.  The overlapping of communities, in a broader sense, should allow communities to share a finite part of their components, consisting of both nodes and edges. This view of the overlappedness of communities is related to the recent discussion of ``dominant communities'' versus ``hidden communities'' \citep{Het2018}, which emphasizes that detected communities are not of the same significance to the graph and may demonstrate different strength, essentially embodying the idea of hierarchical community structures (see below).

\subsection*{Seed Set Expansion}

Alongside the abandoning of hard-partitions, which inspired a lot of metric-based optimization methods that directly deal with the entire graph, people gradually adopted the new belief that \textit{locality} matters in the determination of communities. In particular, a local determination is more consistent with the logic behind the formulation of communities in real social networks, where nodes often do not have a clear sense of the entire network and groups mostly emerge from local commonalities. Adopting this modern view, a new category of algorithms for community detection, termed as the seed set expansion process, has been gaining more and more attention. The idea is to start with finite seed sets and expand them into communities by adding/removing nodes to/from the set if a certain measure of the community is improved, such as modularity \citep{C2005}, conductance \citep{AL2006}, outwardness \citep{B2008}, fitness \citep{LLet2009}, or significance \citep[OSLOM,][]{LLet2011}. One important line of seed set expansion algorithms originate from the PageRank algorithm and expand the seed set based on a random walk process, as pioneered by the work of \citet{AL2006} and \citet{Aet2006}. \citet{Let2013} proposed an algorithm in which the seed set is determined based on clique-detection methods, as cliques could essentially be viewed as communities cores \citep{Pet2005, Ret2012}. \citet{KK2014} studied different seed set expansion algorithms through a comparative analysis, focusing on the determination of a good seed set. More recently, \citet{Get2016} proposed a core identification strategy, an algorithm based on the DBSCAN method \citep{Eet1996,Cet2013} where two parameters are adopted: (1) $\epsilon$ defines the radius of the neighborhood of a node that is considered; (2) $MinPts$ is the minimum number of neighbors of a node's $\epsilon$-neighborhood; nodes are defined as cores if they have more than $MinPts$ neighbors in their $\epsilon$-neighborhood. Similarly, \citet{Bet2017} proposed an algorithm for overlapping community detection using the nodes that are density peaks as community cores, an idea borrowed from clustering analysis \citep{RL2014}. Nodes with high local density $\rho$ and large distance $\delta$ from other density peaks are identified in the $\rho-\delta$ plot as community cores, around which other nodes are classified.

We notice that, among existing seed sed expansion methods, a few problems arise. First, many existing algorithms make ad-hoc decisions on the seed set or the community core (e.g. cliques), which often consists of an arbitrary number of nodes. Clique percolation methods use cliques as the seed sets, while the size of the cliques is experimentally decided \citep{Xet2013}. \citet{KK2014} shows that in fact a random seed set may yield better performance than a seed set selecting high-degree nodes. \citet{LLet2009} invented the notion of the ``natural community" of nodes, which essentially serves as the community cores. There is little agreement on how many nodes a seed set should consist, and what is the order for these seeds to join the set, if the set has multiple nodes. We argue that it is more natural to assume that in most cases initially each seed set only contains a single node, and all other nodes sequentially joining the set should follow a hierarchical order; only for the rare case that neighboring nodes have completely identical topological features, could a seed set consists more than one node. The second problem is that in most existing algorithms, the expansion process is in fact still non-local: it does not allow each node \textit{itself} to decide whether it should join a community, and in many cases the stopping criterion for expansion is still from an optimization standpoint. As we mentioned, in social network settings, nodes themselves are often ignorant about the nature of the entire network, which leads to the idea of seed set expansion; moreover, most likely nodes are also unaware of the situation of the rest of their belonged communities: they don't know if their joining or leaving the group will maximize some metrics of the community, and even they do, this may not be the factor that influences their decision. Therefore, we believe that the stopping criterion for seed set expansion is supposed to follow a more behavioral rule when dealing with human networks.

\subsection*{Label Propagation}

The second problem for the abovementioned seed set expansion algorithms, that they assume subsequent nodes are attached to the communities in a static and non-local fashion, could be resolved by an advanced idea, that the community assignment of non-core nodes is determined from a propagatory standpoint. This brings in the idea of another line of community detection methods, known as label propagation algorithms, first proposed by \citet{Ret2007} and having seen a lot of variants thereafter (for example, the speaker-listener SLPA \citep{Xet2011}, the DEMON algorithm \citep{Cet2012} and more recent designs \citep[e.g.][]{CR2017}). The idea of label propagation is simple: iteratively each node sends the label of its community membership to its neighbors, and at each time step the node's community membership is updated based on the information it receives from all neighbors, according to certain decision rules (e.g., a majority vote \citep{Ret2007} or a listener-speaker scenario \citep{Xet2011}); and eventually, the algorithm will stop at convergence, i.e., there is no more update of community membership on any node during the propagation.

We believe that this dynamic and propagatory point of view for community detection is important in social network settings: nodes compete with each other trying to expand their influence, and finally the winners will be able to establish their communities. It is more advanced than the traditional view that community membership is a priori determined from a global optimization standpoint. We agree that community assignments should definitely rely on the graph's topology, but instead of regressing the community membership to simplified metrics of the topology, it may be more organic (especially for social networks) to set up the propagation and let the dynamics decide the equilibrium convergence. By this means, label propagation algorithms successfully highlight \textit{complete} locality in the determination of community memberships, as no optimization at any non-individual level is assumed. However, one significant problem for this approach is that the propagation could follow arbitrary rules, and thus each proposal of a different rule for community decisions will possibly end up with a new algorithm, which suggests that the label propagation idea essentially consists of an unlimited algorithmic space. Inevitably, this triggers debates on a good decision rule that processes a node's information received from different neighbors. Moreover, decision rules in the first generation of label propagation algorithms often select one community label for each node from all candidates and thus result in hard-partitioning; to apply label propagation in overlapping community detection, improved designs are to be invented.

\subsection*{Nodes with Memory}

The above difficulty could be overcome by a new generation of label propagation algorithms that introduce a finite memory associated with each node. With the memory kernel storing the information during the propagation (infection) process, detection algorithms are now able to carry out overlapping communities results \citep{G2010}. The idea of nodes with memories is aligned with the term ``fuzzy detection'' \citep{Xet2011}; it retains more information of the propagation process than simplistic decision rules leading to hard-partitioning (e.g. the majority vote), although in existing designs some infection information is still compromised \citep{G2010}, such as the receiving order of labels. From the node's memory, the finite infection history it experienced could be revealed and then used to decide its multiple community membership. Given these considerations, we argue that the memory of nodes is an important feature for effective overlapping community detection methods based on the propagation process. Moreover, as a side note, another problem of the algorithm in \citet{G2010} is that it requires a pre-determined number of communities in the graph in order to set the dimension of the memory vector, which we believe is not necessary. 

\subsection*{Multi-step Detection and Hierarchical Structures}

While seed set expansion based on label propagation process is a modern and arguably successful heuristic for community detection, one should note that a complete seed set expansion scheme is multi-step, and it requires specific algorithm design for each step of the workflow. Unfortunately, most previous studies focused on one stage of seed set expansion and few efforts have been made on designing the workflow of the expansion process.

\citet{Let2008} proposed a multi-step community discovery scheme for textual data where each node is a piece of text. First, the seeding cores are identified using the \textit{Apriori} algorithm; then the detected cores are merged based on similarity; after the determination of cores, all other nodes are assigned to communities relying on their connectivity conditions; and finally, a classification step is applied to make sure that each node belongs to the right community and false assignments are removed. \citet{Wet2013} proposed another multi-step detection algorithm based on seed set expansion. The algorithm consists of four stages: filtering, seeding, seed set expansion and propagation. At the first stage the graph is pruned to core components that are densely connected, and the peripheral structures are omitted. The seed set is determined in the next stage, around which communities are formulated, using the spectral method based on the optimization of conductance, originated from \citet{Aet2006}. The omitted peripheral structures are reinstalled to the detected communities at the final stage. 

Multi-step algorithms extensively appear in the detection of hierarchical community structures \citep[e.g.][]{Cet2008, LLet2009, LLet2011}, which has been drawing more and more attention recently. The idea of hierarchical communities is that the detection of communities should associate the partitioning with an order of significance, possibly through a hierarchy, instead of treating all detected communities equally, as most existing methods do. \citet{Set2007} proposed a method uncovering the hierarchical organizations of nodes based on a new node-affinity metric and on searching for the local maxima of modularity. \citet{Set2008} designed a multi-step algorithm named EAGLE to detect hierarchical and overlapping community structures, where maximal cliques in the graph are used as the seed set and an agglomerative process relying on modularity maximization helps establish the hierarchy. A similar multi-step algorithm named SHRINK was proposed by \citet{Set2010}, where each node is assigned with an initial label and the (multi-ary, as opposed to binary) hierarchical community structure is gradually established by measuring the modularity gain of merging end-communities. \citet{P2014} studied the hierarchical structure of the SBM and proposed an inference algorithm to select the best multi-level hierarchical model, which facilitates the formation of benchmark hierarchical SBM graphs for testing detection algorithms. Recently, a recursive bi-partitioning algorithm is devised with a top-down partition workflow \citep{LLet2018}, as opposed to the agglomerative process \citep[e.g.][]{Set2008, Set2010}. Overall, multi-step algorithms based on certain propagation processes that consider the hierarchical structure of communities, as emerged in this evolutionary discussion, may contribute the modernest ideas to current community detection methodologies.

\subsection*{Isolated Nodes}

As a last note, it should be pointed out that the attention to the peripheral structures of the graph, besides the densely connected cores, is non-trivial, which is relevant to the idea of ungrouped isolated nodes \citep{Wet2013}. The belief is that, not all nodes belong to communities; isolated nodes (noises) do exist. \citet{Get2005} regarded them as ``unstable nodes'' and discussed the determination of these nodes through essentially a Monte-Carlo approach by imposing random noises on edge weights. \citet{Get2018} proposed a seed-set-based label propagation algorithm that discovers ``boundary nodes'' as opposed to ``core nodes'', whose basic idea is similar. \citet{Set2010} also discussed the ``hubs'' and ``outliers'' among ``homeless'' nodes identified in the detection process. Nevertheless, in general the notion of isolated nodes is often neglected by existing works and people tend to assign community memberships to all nodes in the graph. 

\section*{Appendix B: Running Time Analysis}

Comparing the running time of our algorithm on various networks, it shows that random networks could possibly be spotted during the label propagation process. The cumulative running time curve for a random network generally does not follow an S-shape, as the case on real networks, and instead demonstrates a more gradual growth (Figure S1, left). This is because on real networks the depth of propagation, i.e., the reachability of end-communities (hubs) is often heterogeneously distributed with a small tail, while on random networks all end-communities tend to have the same topological features and thus the simultaneous propagation from different hubs is more gradual and synchronized. However, this criterion may lead to a wrong catch since the curves for some real (traditional) networks are also not well S-shaped. 

It shows that the total running time of the detection scheme has a quasi-linear relationship with the network size (Figure S1, right), demonstrating the $O(E)$ time complexity of our algorithm (equation (5)).

\renewcommand{\thefigure}{S\arabic{figure}}
\setcounter{figure}{0}

\begin{figure*}
\begin{minipage}{\linewidth}
\centering   
	\includegraphics[width=6in]{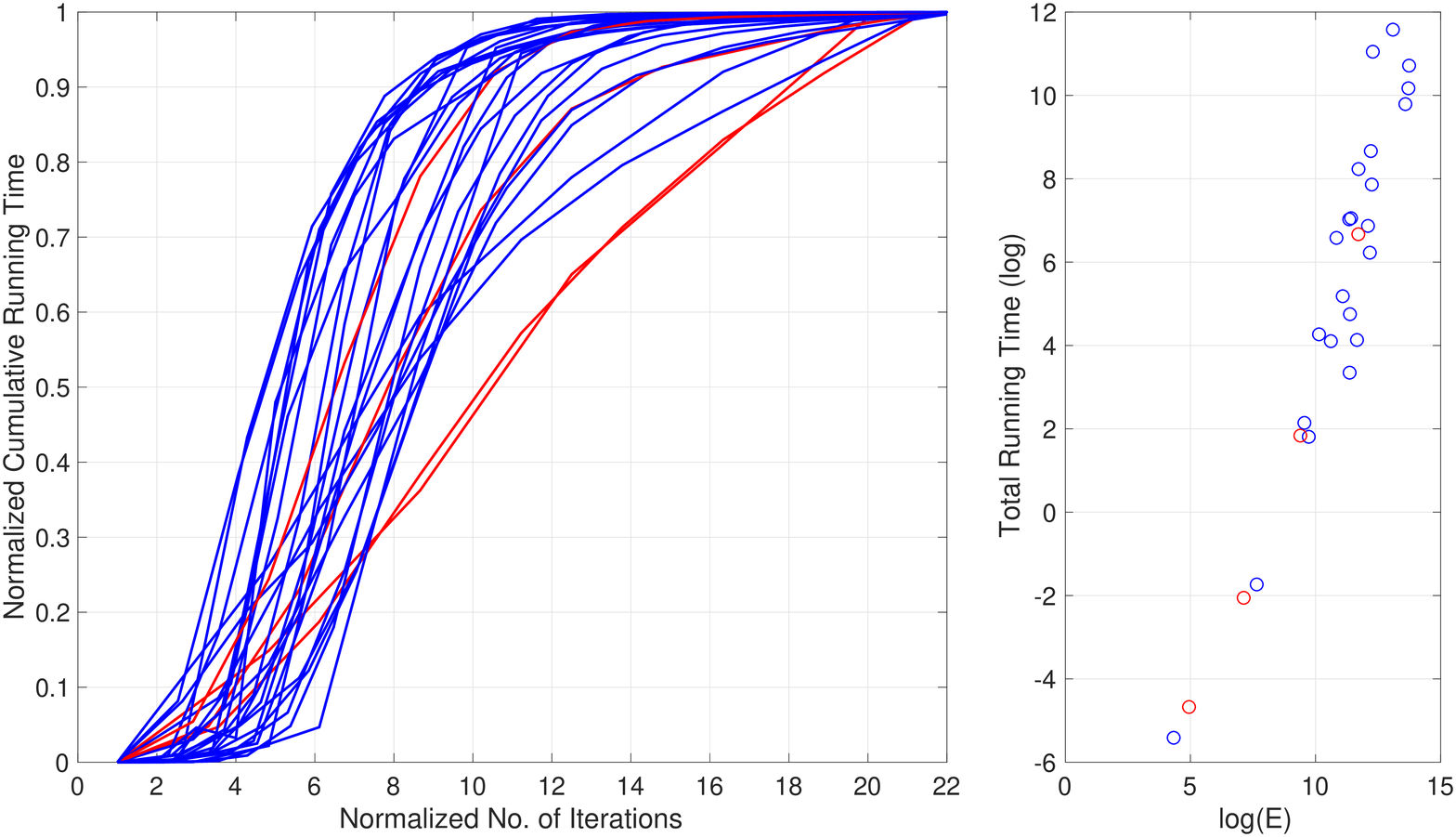}  
\caption{Running time of our detection scheme on various networks. Random networks shown in red; real networks shown in blue. Left: cumulative running time at Step 2 as a function of the number of iterations ($t_{fin}$). Both axis normalized to scale to account for various lengths. Right: total running time (in seconds) as a function of the number of edges (log-log plot). The $O(E)$ time complexity of the algorithm could be identified.} 

\end{minipage}
\end{figure*}

For other detection methods, tests show that SLPA and DEMON demonstrate a good $O(E)$ time complexity, similar to our algorithm, while the running time of HLC or Perco does not exhibit a consistent dependence on the scale of the network (Figure S2). As is acknowledged, a highly variant computational time undermines the applicability of the detection algorithm.

\begin{figure*}
\begin{minipage}{\linewidth}
\centering   
	\includegraphics[width=6in]{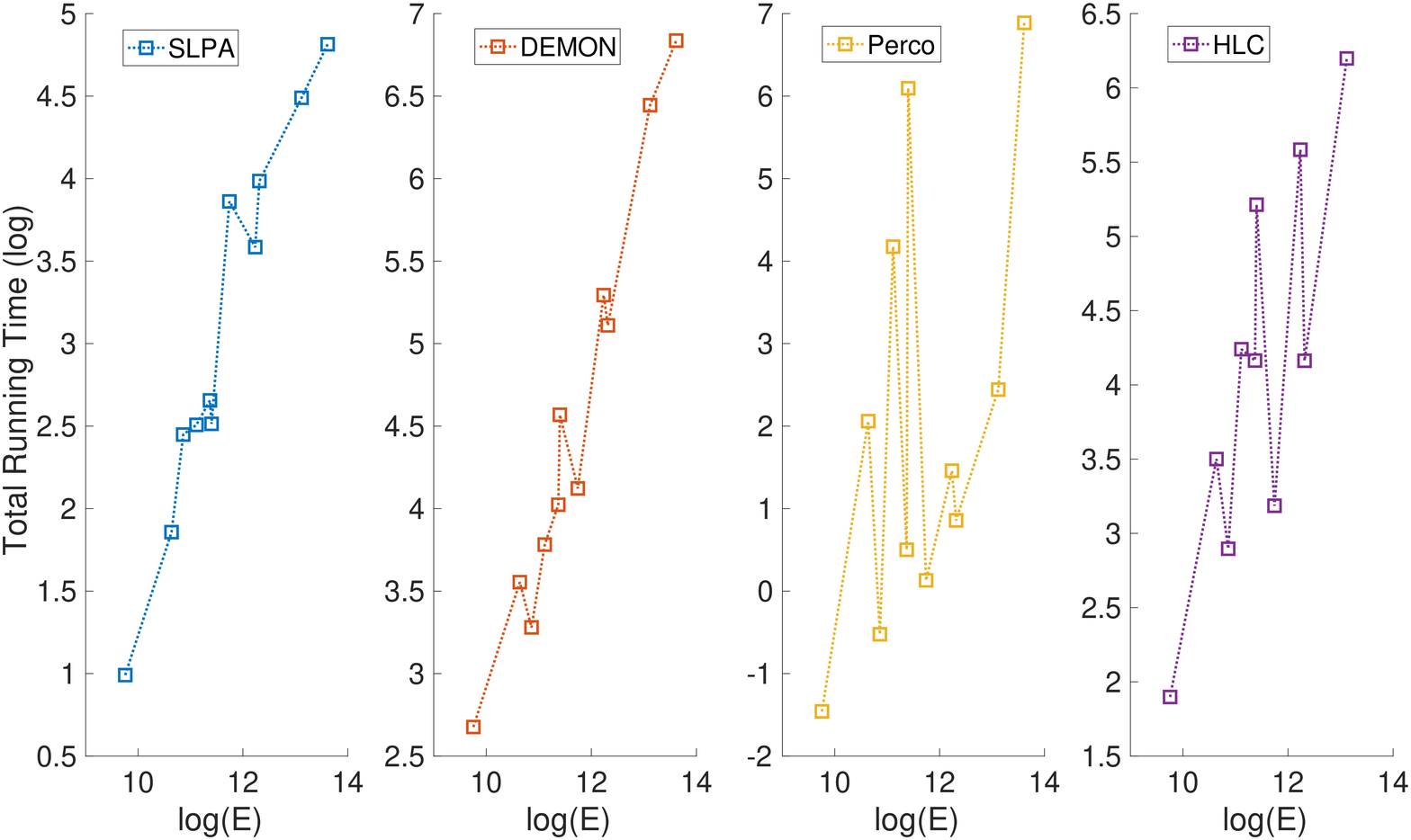}  
\caption{Running time of reference detection algorithms. In general, SLPA and DEMON exhibit $O(E)$ time complexity; HLC and Perco do not.} 

\end{minipage}
\end{figure*}

\section*{Appendix C: Cutoff Level of the Community Hierarchy}

The determination of the cutoff level of the community hierarchy on large real networks could refer to the $\epsilon \leftrightarrow \Delta |R_{\epsilon}|$ and $\epsilon\leftrightarrow\Phi_{\epsilon}$ curves (Figure S3), following the same heuristics explained on the LFR network (Figure 4). The local peaks on these two curves indicate potential cutoff levels of $\epsilon$. In practice, two peaks often agree on the same value, which implies a good cutoff.

\begin{figure*}
\begin{minipage}{\linewidth}
\centering   
	\includegraphics[width=6in]{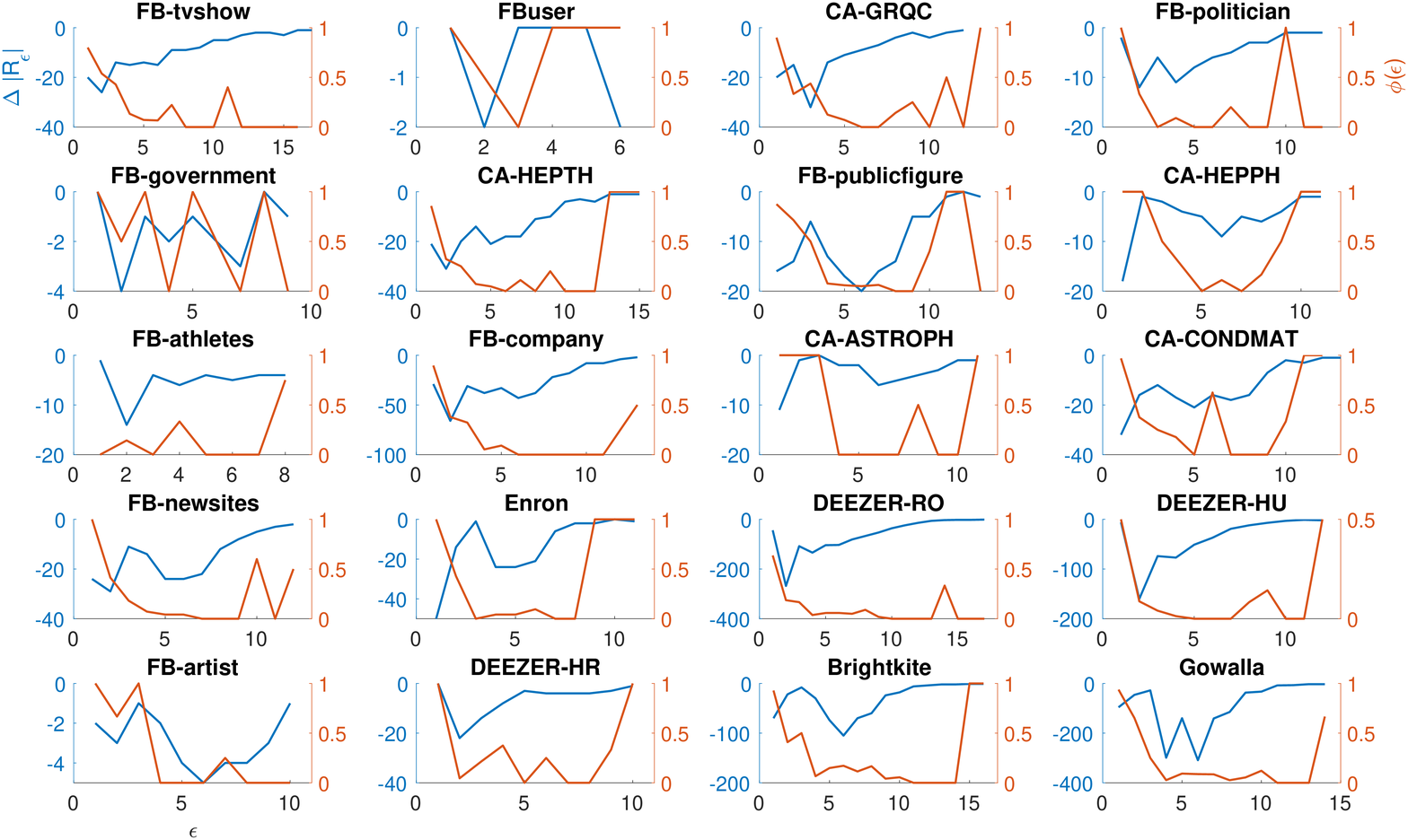}  
\caption{$\epsilon \leftrightarrow \Delta |R_{\epsilon}|$ (blue) and $\epsilon\leftrightarrow\Phi_{\epsilon}$ (red) curves for tested networks. Local peaks on the two curves suggest possible cutoffs.} 
\end{minipage}
\end{figure*}

\section*{Appendix D: Performance of Different Detection Algorithms on the Karate club network}

Different detection algorithms are tested on the Karate club network (Figure S4). The SLPA algorithm is not deterministic even on the small-scale network, and one detection result is shown. HLC detected 20 communities from 34 nodes and 78 edges, which is clearly not satisfactory and hence the results are not shown. DEMON and Perco yielded deterministic and meaningful results. Compared with other algorithms, our detection scheme won by a large margin on the Karate club network; the results are reliable and match the ground truth to a great extent (Figure 2).

\begin{figure*}
\begin{minipage}{\linewidth}
\centering   
	\includegraphics[width=5in]{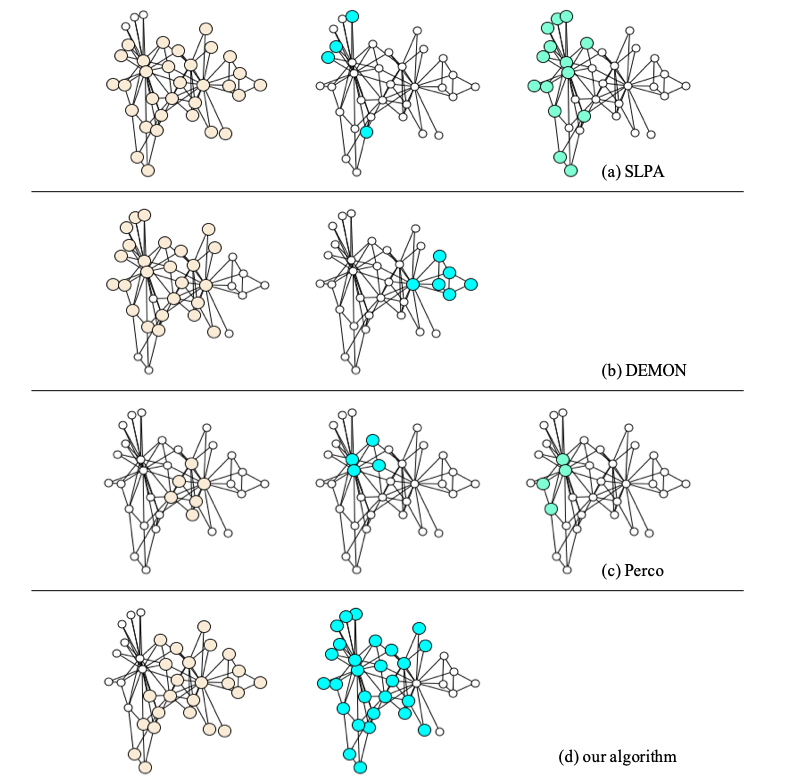}  
\caption{Community detection results of SLPA, DEMON, Perco and our algorithm on the Karate club network. HLC detected 20 communities on the network and the results are not shown.} 
\end{minipage}
\end{figure*}

\end{document}